\documentclass[journal,final]{IEEEtran}
\usepackage[T1]{fontenc}
\usepackage{bm}
\usepackage{amsmath}
\usepackage{amsthm}
\usepackage{amssymb}
\usepackage{graphicx}

\usepackage{stfloats}
\usepackage{subfigure} 
\usepackage{booktabs}
\usepackage{multicol}
\usepackage{multirow}
\usepackage{makecell}
\usepackage{color}

\date{}

\usepackage[unicode=true,
bookmarks=true,bookmarksnumbered=true,bookmarksopen=true,bookmarksopenlevel=1,
breaklinks=false,pdfborder={0 0 0},pdfborderstyle={},backref=false,colorlinks=false]
{hyperref}
\hypersetup{pdftitle={Your Title},
	pdfauthor={Your Name},
	pdfpagelayout=OneColumn, pdfnewwindow=true, pdfstartview=XYZ, plainpages=false}

\makeatletter
\theoremstyle{plain}
\newtheorem{thm}{\protect\theoremname}
\theoremstyle{definition}
\newtheorem{defn}[thm]{\protect\definitionname}
\theoremstyle{remark}
\newtheorem{rem}[thm]{\protect\remarkname}
\theoremstyle{plain}
\newtheorem{lem}[thm]{\protect\lemmaname}
\theoremstyle{plain}
\newtheorem{prop}[thm]{\protect\propositionname}

\usepackage[vlined,boxed,ruled,linesnumbered,resetcount]{algorithm2e}
\usepackage{algpseudocode}
\usepackage[noadjust,sort,compress]{cite}

\providecommand{\definitionname}{Definition}
\providecommand{\lemmaname}{Lemma}
\providecommand{\propositionname}{Proposition}
\providecommand{\remarkname}{Remark}
\providecommand{\theoremname}{Theorem}
\usepackage[justification=centering]{caption}

\newcommand {\aplt} {\ {\raise-.5ex\hbox{$\buildrel<\over{\mbox{\scriptsize $\sim$}}$}}\ }

\renewcommand{\IEEEQED}{\IEEEQEDopen}

\usepackage{color, colortbl}
	
\definecolor{Gray}{gray}{0.9}
\usepackage[first=0,last=9]{lcg}

\makeatother
 
\begin{document}
\title{Better Lattice Quantizers Constructed from Complex Integers}

\author{Shanxiang Lyu, Zheng Wang, Cong Ling, Hao Chen
			\thanks{This work was supported in part by the National Natural Science Foundation of China under Grants 61902149, 61801216, and 62032009,   the Natural Science Foundation of Guangdong Province under Grant 2020A1515010393, and  the Major Program of Guangdong Basic and Applied Research under Grant 2019B030302008.
			}
	\thanks{
	Shanxiang Lyu and Hao Chen are with the College of Cyber Security, Jinan
		University, Guangzhou 510632, China (Email: shanxianglyu@gmail.com, chenhao@fudan.edu.cn), Zheng Wang is with the School of Information Science and Engineering, Southeast University, Nanjing 210096, China (Email: z.wang@ieee.org),
	 Cong Ling is with the Department of Electrical and Electronic Engineering,
	 Imperial College London,
	 London, SW7 2AZ, United Kingdom (Email: c.ling@imperial.ac.uk).
	}
}
\maketitle

\begin{abstract}
	This paper investigates low-dimensional quantizers from the perspective of complex lattices.
	We adopt Eisenstein integers and Gaussian integers to define checkerboard lattices $\mathcal{E}_{m}$ and $\mathcal{G}_{m}$. By explicitly linking their lattice bases to various forms of $\mathcal{E}_{m}$ and $\mathcal{G}_{m}$ cosets, we discover the $\mathcal{E}_{m,2}^+$ lattices, based on which we report the  best known lattice quantizers in dimensions $14$, $15$, $18$, $19$, $22$ and $23$.
	 Fast quantization algorithms of the generalized checkerboard lattices are proposed to enable evaluating the normalized second moment (NSM) through Monte Carlo integration. 
\end{abstract}

\section{Introduction}
\noindent
 \IEEEPARstart{T}{he} theory of lattices has been used to  achieve remarkable
 breakthroughs in a wide range of fields, ranging from communications to cryptography.
 Most of the applications require the construction of good lattices for sphere-packing (e.g., channel coding \cite{tit/Forney88a,Erez2004,tit/CampelloLB18}) or quantization (e.g., lossy source coding \cite{zamir2014lattice}, 
 spatial lattice modulation \cite{DBLP:journals/tsp/ChoiNL18}, data hiding \cite{lin2021lattice}). For a long time sphere-packing has attracted the attention of Mathematicians. Many low-dimensional dense lattices have been found \cite{Conway1999}, whose optimality have been proved in dimensions $3$, $8$, and $24$ \cite{hales2005proof,viazovska2017sphere,cohn2017sphere}.
 

Compared to sphere-packing,
optimal lattices for quantization are less developed.
The optimal lattice quantizer refers to the lattice that features the smallest normalized second moment (NSM).
If the lattice is used as a quantizer, all the points in the Voronoi region around the lattice point $\mathbf{y}$ are represented by $\mathbf{y}$.
In dimensions $13\leq n \leq 23$, most of the best known lattice quantizers
 have been   $D_n^*$ and $A_n^*$ \cite{allen2021optimal} (except  
the Barnes–Wall lattice $\Lambda_{16}$
and the 
tailbiting codes based lattice in $n=20$ \cite{isit/KudryashovY10}), whose NSMs are much larger than Zador's upper bound. 
%
Recently Agrell and Allen \cite{agrell2022}   employed the technique of  product lattices to improve lattice quantizers in these dimensions, but the quantizers may still be far from being optimal (see \cite[Thm. 7]{agrell2022}).

 
To construct a lattice quantizer, it seems more rewarding to start from an algebraic approach \cite{Conway1999} rather than a random-search approach \cite{agrell98}. Many known optimal lattices exhibit  a high degree of symmetry, which can be induced by constructing algebraic lattices through rings of number fields. Compared to high order cyclotomic fields, quadratic fields and complex integers are conceptually simpler.
By using complex Constructions A and B to lift linear codes to lattices, Conway and Sloane \cite[Chap. 7]{Conway1999} have shown that many optimal low-dimensional lattices can be produced.
In addition,
 algebraic lattices often enjoy faster quantization/decoding algorithms. E.g., complex lattices defined by Gaussian integers and Eisenstein integers have been used to construct lattice reduction algorithms which are about $50\%$ faster than their counterparts \cite{lyu20im,Gan2009}. 
 
 This paper attempts to further advance low-dimensional lattice quantizers from the perspective of complex lattices. The contributions are summarized as follows:

\begin{itemize}
	\item  We discover the $\mathcal{E}_{m,2}^+$ lattices, which exhibit the best reported NSM in dimensions $14$, $18$, and $22$. These lattices are built by appropriately choosing the union of cosets from the complex-valued checkerboard lattices, where the crux is to link the lattice bases to various forms of cosets.  The product lattices based on $\mathcal{E}_{m,2}^+$  also achieve the best reported quantizers in dimensions  $15$, $19$, and $23$.
The generalized checkerboard lattices from the perspectives of Eisenstein integers and Gaussian integers include the equivalent forms of the celebrated $D_4$, $E_6^*$ and $E_8$ lattices. 
 In the context of applications where low-dimensional lattice quantizers are popular (see \cite{DBLP:journals/tsp/ChoiNL18,lin2021lattice}), the proposed quantizers can be employed to achieve the smallest NSM in their respective dimensions.

	\item  We  present efficient quantization algorithms for the proposed generalized checkerboard lattices, which are denoted as 
	$Q_{\mathcal{E}_{m}}$, $Q_{\mathcal{G}_{m}}$, $Q_{\mathcal{E}_{m,2}^+}$, $Q_{\mathcal{E}_{m,1+\omega}^+}$, $Q_{\mathcal{G}_{m,2}^+}$, and $Q_{\mathcal{G}_{m,1+i}^+}$. The rationale behind 	$Q_{\mathcal{E}_{m}}$ and $Q_{\mathcal{G}_{m}}$ is to modify the coefficients after component-wise quantization, whereas the principle of $Q_{\mathcal{E}_{m,2}^+}$, $Q_{\mathcal{E}_{m,1+\omega}^+}$, $Q_{\mathcal{G}_{m,2}^+}$, and $Q_{\mathcal{G}_{m,1+i}^+}$ is to use coset decomposition. With the aid of the presented algorithms, the NSM of the proposed lattices can be numerically evaluated through Monte Carlo integration.  
\end{itemize}
 
  
 Notation: Matrices and column vectors are denoted by uppercase and
 lowercase boldface letters. The sets of all rationals, integers, real and complex numbers
 are denoted by $\mathbb{Q}$, $\mathbb{Z}$, $\mathbb{R}$ and $\mathbb{C}$,
 respectively.   $\oplus$,
 $\underset{K}{\otimes}$ and $\otimes$ denote the direct sum, the Kronecker tensor product and the Cartesian product, respectively. $\mathrm{sum}(\cdot )$ represents the summation of all the components in a vector. $Q_{\mathcal{S}}(\cdot)$ is the nearest neighbor operator that finds the closest element/vector of the set $\mathcal{S}$ to the input.
 $\mathcal{R}(\cdot)$ and $\mathcal{I}(\cdot)$ are the operators of getting the real and imaginary parts of the input, respectively. $\approxeq$ denotes the equivalence of lattices.
 
\section{Preliminaries}
\subsection{Real-valued Lattices}
\begin{defn}[Real lattice]
	An $n$-dimensional lattice $\Lambda$ is a discrete additive subgroup of $\mathbb{R}^{n'}$, $n' \geq n$.
Consider $n$ linearly independent vectors $\mathbf{b}_{1},\ldots\thinspace,\mathbf{b}_{n}$ in $\mathbb{R}^{n}$,  the associated lattice is represented by
	\begin{equation}
	\Lambda = \mathcal{L}(\mathbf{B}) =\lbrace z_1\mathbf{b}_{1}+ z_2\mathbf{b}_{2}+ \cdots + z_n\mathbf{b}_{n}:  z_1,...,z_n \in \mathbb{Z}\rbrace.
	\end{equation}
 $\mathbf{B}=[\mathbf{b}_{1},\ldots\thinspace,\mathbf{b}_{n}]$ is referred to as the generator matrix (lattice basis) of $\Lambda$. 
\end{defn}

``Quantization'' denotes the map
from a vector  $\mathbf{y}\in \mathbb{R}^n$ to the closest lattice point of $\Lambda$:
\begin{equation}\label{eq_quandef}
Q_\Lambda(\mathbf{y})=\mathop{\arg\min}_{\bm{\lambda} \in \Lambda} \|\mathbf{y}-\bm{\lambda}\|.
\end{equation}
The r.h.s. of Eq. (\ref{eq_quandef}) is known as solving the closet vector problem (CVP) \cite{Micciancio2002} of $\Lambda$, which requires efficient algorithms to do so. The CVP of $\Lambda$ can be adapted to its coset $\mathbf{g}+\Lambda$:
\begin{equation}\label{cosetcvp_def}
Q_{\Lambda+\mathbf{g}}(\mathbf{y}) = \mathbf{g} + 	Q_{\Lambda}( \mathbf{y}-\mathbf{g}).
\end{equation}

 
The   Voronoi region $\mathcal{V}_\Lambda$ of a lattice $\Lambda$   is the convex polytope
\begin{align}
\mathcal{V}_{\Lambda}=\{\mathbf{y} \in \mathbb{R}^n: \left\|\mathbf{y}\right\|^2 \leq \left\|\mathbf{y}-\bm{\lambda}\right\|^2 \text{ for all } \bm{\lambda} \in \Lambda\}.
\end{align}
Since $\mathbf{y}-Q_\Lambda(\mathbf{y})\in \mathcal{V}_{\Lambda}$, the quantizer's properties are determined by $\mathcal{V}_{\Lambda}$. The NSM  of a lattice $\Lambda$ is defined as
\begin{equation}\label{defNSM}
G_{n}(\Lambda)= \frac{\int_{\mathbf{x}\in \mathcal{V}_{\Lambda} } ||\mathbf{x}||^2 \mathrm{d}\mathbf{x}}{n \rm{Vol}({\Lambda})^{1+\frac{2}{n}}},
\end{equation}
where $ \rm{Vol}({\Lambda})=\rm{det}(\mathbf{B}^\top \mathbf{B})^{1/2}$ is referred to as the volume of $\Lambda$.

\subsection{Complex-valued Lattices}
\begin{defn}[Quadratic field]
	A quadratic field is an algebraic number field $\mathbb{K}$ of degree
	$[\mathbb{K}:\mathbb{Q}]=2$ over $\mathbb{Q}$.  For a square free positive integer $d$, we say $\mathbb{K}=\mathbb{Q}\left(\sqrt{-d}\right)$
	is an imaginary quadratic field.
\end{defn}

\begin{defn}[Complex integer]
	The set of algebraic integers in $\mathbb{Q}\left(\sqrt{-d}\right)$ forms a ring of integers denoted as $\mathbb{Z}\left[\xi\right]$, where $\xi=\sqrt{-d}$ if $-d \equiv 2,3 \mod 4$, and $\xi=(1+\sqrt{-d})/2$ if $-d \equiv 1 \mod 4$.
\end{defn}
By setting $d=1$, we obtain the set of Gaussian integers $\mathbb{Z}\left[i\right]$, $i\triangleq \sqrt{-1}$. By setting $d=3$, we obtain the set of Eisenstein integers $\mathbb{Z}\left[\omega\right]$, $\omega \triangleq \frac{1+\sqrt{-3}}{2}$ ($\omega$ is set as the 
sixth root of unity for convenience, rather than the third root of unity).



\begin{defn}[Complex lattice \cite{lyu20im}]
An $m$-dimensional complex lattice $	\bar{\Lambda}$ is a discrete
	$\mathbb{Z}\left[\xi\right]$-submodule of $\mathbb{C}^{m'}$ that has a basis, $m'\geq m$. Consider $m$ linearly independent vectors 	$\bar{\mathbf{b}}_{1},\ldots\thinspace,\bar{\mathbf{b}}_{m}$  in 
	$\mathbb{C}^{m}$,  the associated complex-valued  lattice is represented by
	\begin{equation}
	\bar{\Lambda} = \mathcal{L}(\bar{\mathbf{B}}) =\lbrace \bar{z}_1\bar{\mathbf{b}}_{1}+ \bar{z}_2\bar{\mathbf{b}}_{2}+ \cdots + \bar{z}_m\bar{\mathbf{b}}_{m}: \bar{z}_1,...,\bar{z}_m \in \mathbb{Z}\left[\xi\right] \rbrace.
	\end{equation}
 $\bar{\mathbf{B}}=[\bar{\mathbf{b}}_{1},\ldots\thinspace,\bar{\mathbf{b}}_{m}]$ is referred to as the generator matrix  of $\bar{\Lambda}$. 
\end{defn}

The complex quantizer is defined as
\begin{equation}\label{eq_complex_quan}
Q_{\bar{\Lambda}}(\bar{\mathbf{y}})=\mathop{\arg\min}_{\bar{\bm{\lambda}} \in \bar{\Lambda}} \|\bar{\mathbf{y}}- \bar{\bm{\lambda}}\|,
\end{equation}
which returns the closest vector to  $\bar{\mathbf{y}}\in \mathbb{C}^m$ over $\bar{\Lambda}$. The r.h.s. of (\ref{eq_complex_quan}) is referred to as the CVP of a complex lattice.

Based on the complex-to-real transform of $\Psi: \mathbb{C}^m \rightarrow \mathbb{R}^{2m}$,
\begin{equation}\label{eq_rct}
	[\bar{x}_1, ..., \bar{x}_m]^\top \rightarrow 	[\mathcal{R}(\bar{x}_1), ..., \mathcal{R}(\bar{x}_m), \mathcal{I}(\bar{x}_1), ..., \mathcal{I}(\bar{x}_m)]^\top,
\end{equation}
$Q_{\bar{\Lambda}}(\bar{\mathbf{y}})$ amounts to a real-valued quantizer $Q_{\Psi(\bar{\Lambda})}(\Psi(\bar{\mathbf{y}}))$. The $2m$-dimensional real-valued lattice $\Psi(\bar{\Lambda})$  has a basis
	\begin{equation}\label{eq:zizlRelation}
	\mathbf{B}_{\Psi(\bar{\Lambda})}=\left[\begin{array}{cc}
	\mathfrak{R}\left(\bar{\mathbf{B}}\right) & -\mathfrak{I}\left(\bar{\mathbf{B}}\right)\\
	\mathfrak{I}\left(\bar{\mathbf{B}}\right) & \mathfrak{R}\left(\bar{\mathbf{B}}\right)
	\end{array}\right]\left(\Phi^{\mathbb{Z}\left[\xi\right]} \underset{K}{\otimes} \mathbf{I}_{m}\right),
	\end{equation}
	where $	\Phi^{\mathbb{Z}\left[\xi\right]}$ denotes the real-valued basis of $\mathbb{Z}\left[\xi\right]$. In particular
	\begin{align}
		&\Phi^{\mathbb{Z}\left[i\right]} = \left[\begin{array}{cc}
		1 & 0\\
		0 & 1
		\end{array}\right]\\
		& \Phi^{\mathbb{Z}\left[\omega \right]} =\left[\begin{array}{cc}
		1 & 1/2\\
		0 & \sqrt{3}/2
		\end{array}\right].
	\end{align}
	The volume and the NSM of  $\bar{\Lambda}$ can both be defined by $\Psi(\bar{\Lambda})$:
	\begin{align}
&	\mathrm{Vol}\left(\bar{\Lambda} \right) \triangleq \mathrm{Vol}\left( \Psi(\bar{\Lambda})\right) = \left| \det\left(\bar{\mathbf{B}}^\dagger\bar{\mathbf{B}}\right)\right|   \det\left(\Phi^{\mathbb{Z}\left[\xi\right]}\right)^{m}.\label{eq:volume}\\
& G_{m}^{\mathbb{C}}(\bar{\Lambda})\triangleq G_{2m}(\Psi(\bar{\Lambda})).
	\end{align}
 

\section{\label{sec:GoodComplexLattices}Generalization of the Checkerboard lattice}
The checkerboard lattice \cite{Conway1999} is defined as 
	\begin{equation}\label{eq_checkdef}
\mathcal{D}_{n} =\{({x}_1,...,{x}_n)\in \mathbb{Z}^n :{x}_1+\cdots + {x}_n \in 2\mathbb{Z}\},
\end{equation}
while the $D_n^+$ family \cite{Conway1999}  is defined as the union of $\mathcal{D}_{n}$ and its cosets.

\begin{figure}[t!]
	\centering
	\includegraphics[width=0.55\textwidth]{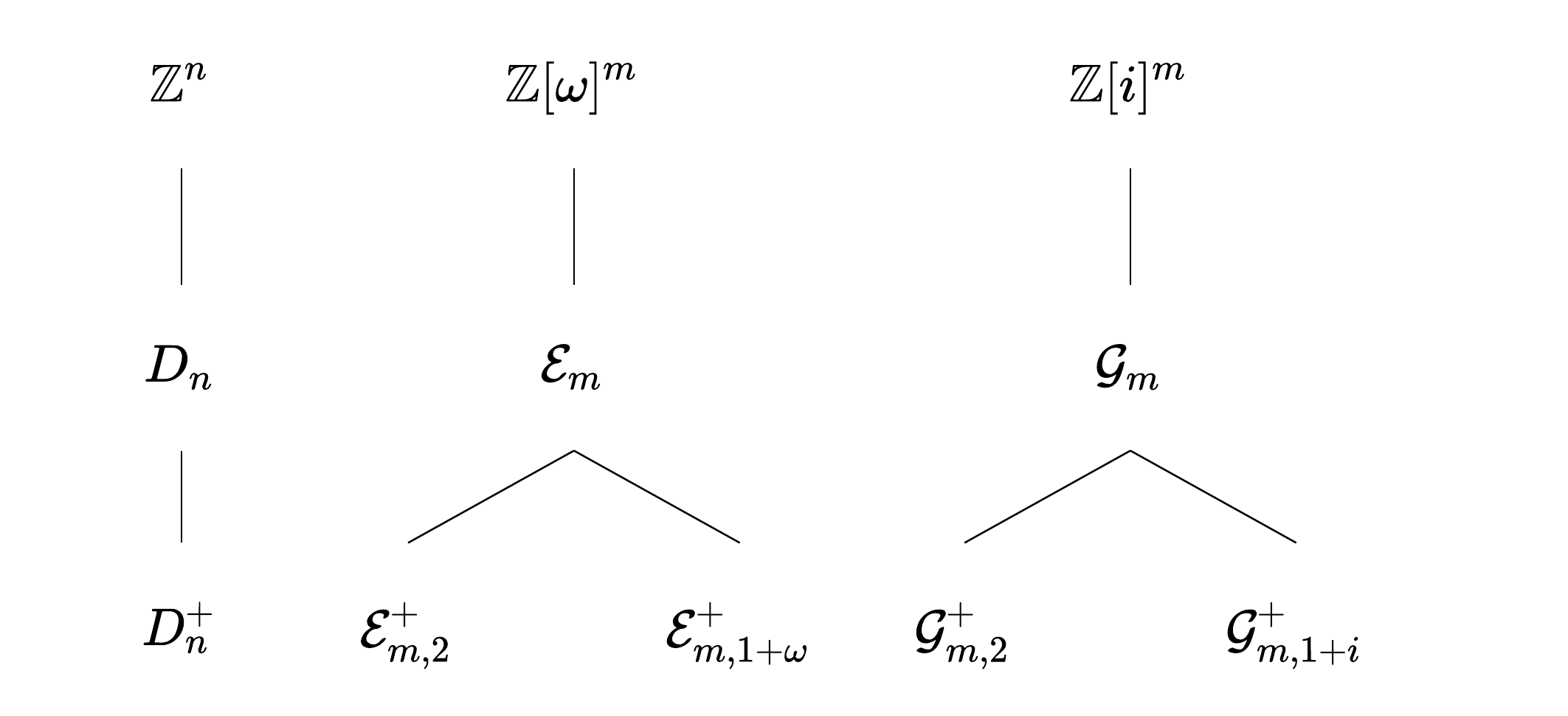}
	\caption{The partition chains of the checkerboard lattice and the generalizations.}
	\label{figPaititon}
\end{figure}

To endow more algebraic properties to $\mathcal{D}_{n}$ and $D_n^+$, we can generalize the real-valued rings to rings of imaginary quadratic fields. The Eisenstein integers $\mathbb{Z}\left[\omega\right]$  and Gaussian integers $\mathbb{Z}\left[i\right]$ have shown promising performance in coding theory (see, e.g., \cite{shum15,Sun2013,jerry2018}), so we adopt such rings to define generalized checkerboard lattices ($\mathbb{Z}\left[\omega\right]$-based   $\mathcal{E}_{m}$, $\mathcal{E}_{m,2}^+$, $\mathcal{E}_{m,1+\omega}^+$, 
and $\mathbb{Z}\left[i\right]$-based $\mathcal{G}_{m}$, $\mathcal{G}_{m,2}^+$, $\mathcal{G}_{m,1+i}^+$).
 The partition chains of these lattices are depicted in Fig. \ref{figPaititon}.
 
%

\subsection{$\mathbb{Z}\left[\omega\right]$-Lattices: $\mathcal{E}_{m}$, $\mathcal{E}_{m,2}^+$ and $\mathcal{E}_{m,1+\omega}^+$}
 
In Eq. (\ref{eq_checkdef}), $2$ is the coefficient with the smallest norm among $\mathbb{Z}$ except $0$ and units. Regarding the coefficient to define the summation of a $\mathbb{Z}\left[\omega\right]$-based checkerboard lattice, it is reasonable to choose $1 + \omega$, which has the smallest norm ($\left\|1 + \omega \right\|=3$)  among $\mathbb{Z}[\omega]$ except $0$ and units \footnote{We have checked other choices for the summation coefficient, but the resulted lattices are inferior to $\mathcal{E}_{m}$.}.
 
\begin{defn}\label{defem}
A sublattice of $\mathbb{Z}[\omega]^m$ is defined as
	\begin{equation}
	\mathcal{E}_{m} =\{(\bar{x}_1,...,\bar{x}_m)\in \mathbb{Z}[\omega]^m :\bar{x}_1+\cdots + \bar{x}_m \in (1 + \omega)\mathbb{Z}[\omega]\}.
	\end{equation}
\end{defn}
 
When $m=1$, the basis of $\mathcal{E}_{m}$ is simply $1 + \omega$. When $m\geq 2$, the following lemma gives the general form of its lattice basis.
\begin{lem}
	If  $\bm{\beta}_1, \bm{\beta}_2\ldots, \bm{\beta}_m \in \mathbb{Z}[\omega]^m$ are $m$ linear independent vectors satisfying
	\begin{enumerate}
		\item $\mathrm{sum}(\bm{\beta}_1) \in (1 + \omega)\mathbb{Z}[\omega], ..., \mathrm{sum}(\bm{\beta}_m) \in (1 + \omega)\mathbb{Z}[\omega]$,
		\item $\left| \det([\bm{\beta}_1, \bm{\beta}_2\ldots, \bm{\beta}_m])\right|^2=3$,
	\end{enumerate}
	then $[\bm{\beta}_1, \bm{\beta}_2\ldots, \bm{\beta}_m]$ is a lattice basis of $\mathcal{E}_{m}$.
\end{lem}
\begin{IEEEproof}
	Condition 1) guarantees that  $[\bm{\beta}_1, \bm{\beta}_2\ldots, \bm{\beta}_m]$  forms either a full lattice or a sublattice of $\mathcal{E}_{m}$.
	Since $\mathbb{Z}[\omega]^m$ consists of 
	\begin{align*}
	&\{(\bar{x}_1,...,\bar{x}_m)\in \mathbb{Z}[\omega]^m :\bar{x}_1+\cdots + \bar{x}_m \in (1+\omega)\mathbb{Z}[\omega]\}\\
	&\cup \{(\bar{x}_1,...,\bar{x}_m)\in \mathbb{Z}[\omega]^m :\bar{x}_1+\cdots + \bar{x}_m \in (1+\omega)\mathbb{Z}[\omega]+1\}\\
	&\cup \{(\bar{x}_1,...,\bar{x}_m)\in \mathbb{Z}[\omega]^m :\bar{x}_1+\cdots + \bar{x}_m \in (1+\omega)\mathbb{Z}[\omega]+\omega\},
	\end{align*}
	we have  $\left| \mathcal{E}_{m}/\mathbb{Z}[\omega]^m \right| =3$.  Thus the condition of $\left| \det([\bm{\beta}_1, \bm{\beta}_2\ldots, \bm{\beta}_m])\right|^2=3$ justifies that $[\bm{\beta}_1, \bm{\beta}_2\ldots, \bm{\beta}_m]$ cannot be a sublattice of $\mathcal{E}_{m}$.
\end{IEEEproof}
Then the lattice basis of $\mathcal{E}_{m}$ can be instantiated as
	\begin{align}
	\bar{\mathbf{B}}_{\mathcal{E}_{m}} = \left[\begin{array}{cc}
\mathbf{I}_{m-1} & \mathbf{0}_{1 \times (m-1)}\\
	\omega \times \mathbf{1}_{1 \times (m-1)} & 1+\omega
	\end{array}\right] \triangleq [\bar{\mathbf{b}}_1, \bar{\mathbf{b}}_2\ldots, \bar{\mathbf{b}}_m],
	\end{align}
where $\mathbf{I}_{m-1}$, $\mathbf{0}_{1 \times (m-1)}$, $\mathbf{1}_{1 \times (m-1)}$ denote  an identity matrix, a column vector of zeros, and a row vector of ones. The subscripts indicate their dimensions.  

\begin{defn}\label{defemEm2}
	The union of $\mathcal{E}_{m}$-cosets is defined as
\begin{equation}\label{EnDecompose23}
\mathcal{E}_{m,2}^+= \mathcal{E}_{m} \cup (\mathcal{E}_{m}+\frac{\bar{\mathbf{d}}}{2})\cup (\mathcal{E}_{m}+\omega\frac{\bar{\mathbf{d}}}{2})\cup (\mathcal{E}_{m}+\omega^*\frac{\bar{\mathbf{d}}}{2}),
\end{equation}
where 
$\bar{\mathbf{d}}=\sum_{k=1}^{m}\bar{\mathbf{b}}_k=[1,1,\ldots,{m\omega}+1]^\top$.
\end{defn}
 
\begin{thm} $\mathcal{E}_{m,2}^+ = \mathbb{Z}[\omega]\bar{\mathbf{d}}/2 \oplus 
	\mathbb{Z}[\omega]\bar{\mathbf{b}}_2 \oplus\cdots \oplus \mathbb{Z}[\omega]\bar{\mathbf{b}}_m$ is a lattice.
%
\end{thm}
\begin{IEEEproof}
By using the decomposition of $\mathbb{Z}[\omega]$ w.r.t. $2\mathbb{Z}[\omega]$, 
\begin{equation}
\mathbb{Z}[\omega] = 2\mathbb{Z}[\omega]\cup \left(2\mathbb{Z}[\omega]+1 \right) \cup
\left(2\mathbb{Z}[\omega]+\omega \right)  \cup \left(2\mathbb{Z}[\omega]+\omega^* \right).
\end{equation}
Multiplying both sides with $\bar{\mathbf{d}}/2$ yields
\begin{align}
&\mathbb{Z}[\omega]\bar{\mathbf{d}}/2 = \mathbb{Z}[\omega]\bar{\mathbf{d}} 
\cup \left(\mathbb{Z}[\omega]\bar{\mathbf{d}} +\bar{\mathbf{d}}/2 \right) \cup \cdots \nonumber\\ 
&\cup
\left(\mathbb{Z}[\omega]\bar{\mathbf{d}} +\omega\bar{\mathbf{d}}/2 \right)  \cup \left(\mathbb{Z}[\omega]\bar{\mathbf{d}} +\omega^* \bar{\mathbf{d}}/2 \right).\label{decompc1}
\end{align}

Recall the definition of $\mathcal{E}_{m}$ is 
\begin{align}
\mathcal{E}_{m} &= \mathbb{Z}[\omega]\bar{\mathbf{b}}_1 \oplus
\mathbb{Z}[\omega]\bar{\mathbf{b}}_2 \oplus \cdots \oplus \mathbb{Z}[\omega]\bar{\mathbf{b}}_m \nonumber \\&= \mathbb{Z}[\omega]\bar{\mathbf{d}} \oplus 
\mathbb{Z}[\omega]\bar{\mathbf{b}}_2 \oplus\cdots \oplus \mathbb{Z}[\omega]\bar{\mathbf{b}}_m. 
\end{align}
By adding $\mathbb{Z}[\omega]\bar{\mathbf{b}}_2 \oplus\cdots \oplus \mathbb{Z}[\omega]\bar{\mathbf{b}}_m$ to both sides of Eq. (\ref{decompc1}), the r.h.s. equals the definition of $\mathcal{E}_{m,2}^+$, while the r.h.s. equals $ \mathbb{Z}[\omega]\bar{\mathbf{d}}/2 \oplus 
\mathbb{Z}[\omega]\bar{\mathbf{b}}_2 \oplus\cdots \oplus \mathbb{Z}[\omega]\bar{\mathbf{b}}_m$.

The independence of $\sum_{k=1}^{m}\bar{\mathbf{b}}_k/2, \bar{\mathbf{b}}_2\ldots, \bar{\mathbf{b}}_m$ follows from the independence of $\bar{\mathbf{b}}_1, \bar{\mathbf{b}}_2\ldots, \bar{\mathbf{b}}_m$. So the $\mathbb{Z}[\omega]$-linear combination of $m$ independent vectors generates a lattice.
\end{IEEEproof}
 
The above theorem immediately shows that $\mathcal{E}_{m,2}^+$ has a lattice basis
\begin{equation}
\bar{\mathbf{B}}_{\mathcal{E}_{m,2}^+} = [\bar{\mathbf{d}}/2,\bar{\mathbf{b}}_2,\ldots,\bar{\mathbf{b}}_m].
\end{equation} 
E.g., the lattice basis of  $\mathcal{E}_{4,2}^+$ can be written as 
\begin{equation}\label{basisE4i}
\bar{\mathbf{B}}_{\mathcal{E}_{4,2}^+}=\left[\begin{array}{cccc}
1/2 & 0 &0 & 0 \\
1/2 & 1 &0 & 0 \\
1/2 & 0 &1 & 0 \\
(4\omega+1)/2 & \omega &\omega & 1+\omega 
\end{array}\right].
\end{equation}

In the same vein, we define
\begin{equation}\label{EnDecompose2}
\mathcal{E}_{m,1+\omega}^+= \mathcal{E}_{m} \cup (\mathcal{E}_{m}+\frac{\bar{\mathbf{d}}}{1+\omega})\cup (\mathcal{E}_{m}+\omega\frac{\bar{\mathbf{d}}}{1+\omega}),
\end{equation} 
which has a lattice basis $\bar{\mathbf{B}}_{\mathcal{E}_{m,1+\omega}^+} = [\bar{\mathbf{d}}/(1+\omega),\bar{\mathbf{b}}_2,\ldots,\bar{\mathbf{b}}_m]$. 

\begin{rem}
	We notice that the $\mathcal{E}_{m}$ lattices have been defined by Jacques Martinet \cite[Section 8.4]{martinet2013perfect}, but the ways we approach them vary significantly. The mathematical treatment on $\mathcal{E}_{m}$ is more thorough in \cite{martinet2013perfect} (e.g., showing whether the lattice is extreme and  eutactic), while we computationally investigate $\mathcal{E}_{m}$ and present the lattice basis. More importantly, the reported better lattice quantizers are due to the novel $\mathcal{E}_{m,2}^+$ lattices we defined, which are generalized from the lattice basis of $\mathcal{E}_{m}$.
	The $\mathcal{E}_{m,1+\omega}^+$ lattices have a similar structure as that of the Coxeter-Todd lattices $K_{2m}$ defined in \cite[Section 8.5]{martinet2013perfect}:
	\begin{align}
	K_{2m} 
	&\approxeq   \mathcal{E}_{m} \cup (\mathcal{E}_{m}+\frac{\mathbf{1}}{\omega^{-1}(1+\omega)})\cup (\mathcal{E}_{m}-\frac{\mathbf{1}}{\omega^{-1}(1+\omega)}).
	\end{align}
Although we fail to find the equivalence between $K_{2m}$ and $\mathcal{E}_{m,1+\omega}^+$, our simulations show that $K_{12}$ and $\mathcal{E}_{6,1+\omega}^+$ exhibit indistinguishable NSM performance (see Section V).
\end{rem}


\subsection{$\mathbb{Z}\left[i\right]$-Lattices: $\mathcal{G}_{m}$, $\mathcal{G}_{m,2}^+$ and $\mathcal{G}_{m,1+i}^+$}
With the aid of Gaussian integers $\mathbb{Z}\left[i\right]$, we  define
\begin{align}
&\mathcal{G}_{m} =\{(\bar{x}_1,...,\bar{x}_m)\in \mathbb{Z}[i]^m :\bar{x}_1+\cdots + \bar{x}_m \in (1+i)\mathbb{Z}[i]\},\\
&\mathcal{G}_{m,2}^+= \mathcal{G}_{m} \cup (\mathcal{G}_{m}+\frac{\bar{\mathbf{p}}}{2})\cup (\mathcal{G}_{m}+i\frac{\bar{\mathbf{p}}}{2})\cup (\mathcal{G}_{m}+(1+i)\frac{\bar{\mathbf{p}}}{2}),\\
&\mathcal{G}_{m,1+i}^+= \mathcal{G}_{m} \cup (\mathcal{G}_{m} +\frac{\bar{\mathbf{p}}}{1+i}),
\end{align}
where 
$\bar{\mathbf{p}}=[1,1,\ldots,mi+1]^\top$. Their lattice bases are:
	\begin{align}
\bar{\mathbf{B}}_{\mathcal{G}_{m}} &= \left[\begin{array}{cc}
\mathbf{I}_{m-1} & \mathbf{0}_{1 \times (m-1)}\\
i \times \mathbf{1}_{1 \times (m-1)} & 1+i
\end{array}\right] \\ 
\bar{\mathbf{B}}_{\mathcal{G}_{m,2}^+} &= \left[\begin{array}{cccc}
1/2 & 0 & \cdots & 0 \\
1/2 & 1 & \cdots & 0 \\
\vdots & \cdots & \ddots & \vdots \\
(mi+1)/2 & i & \cdots & 1+i 
\end{array}\right] \\
\bar{\mathbf{B}}_{\mathcal{G}_{m,1+i}^+} &= \left[\begin{array}{cccc}
1/(1+i) & 0 & \cdots & 0 \\
1/(1+i)  & 1 & \cdots & 0 \\
\vdots & \cdots & \ddots & \vdots \\
(mi+1)/(1+i)  & i & \cdots & 1+i 
\end{array}\right].
\end{align}

 Although these extensions fail to discover better lattice quantizers, they can reproduce some optimal lattices in small dimensions.  
 
 \subsection{Connection to Existing Lattices}
 
 	Two lattices are said to be equivalent if we can obtain one from the other by  scaling, reflection, rotation and unimodular multiplication. For real-valued or complex-valued lattices,  we have $\mathcal{L}(\mathbf{B}_1) \approxeq \mathcal{L}(\mathbf{B}_2)$, $\mathcal{L}(\bar{\mathbf{B}}_1) \approxeq \mathcal{L}(\bar{\mathbf{B}}_2)$  if the lattice bases satisfy
 	\begin{align}
 		\mathbf{B}_1 &=\alpha \mathbf{F} \mathbf{B}_2 \mathbf{U} \label{eqi_eq1} \\
\bar{\mathbf{B}}_1 &=\alpha \bar{\mathbf{F}} \bar{\mathbf{B}}_2 \bar{\mathbf{U}} 
 	\end{align}
 	where $\alpha\in \mathbb{R}$ is a scaling factor, $\mathbf{U}\in \mathrm{GL}_n(\mathbb{Z})$ and $\bar{\mathbf{U}}\in \mathrm{GL}_n(\mathbb{Z}[\xi])$ are unimodular matrices, $\mathbf{F}$ and $\bar{\mathbf{F}}$ are unitary matrices that preserve the (Hermitian) inner product.
 	
 	
 	Then for $	t \in \left\lbrace 0,1,2,...,5\right\rbrace$, by the distance preserving rotation $\omega^t$ we have
 	\begin{align}
 	&\mathcal{E}_{m} \approxeq \omega^t \mathcal{E}_{m} \\
 	&= \{(\bar{x}_1,...,\bar{x}_m)\in \mathbb{Z}[\omega]^m :\bar{x}_1+\cdots + \bar{x}_m \in (1 + \omega)\omega^t\mathbb{Z}[\omega]\}.
 	\end{align}

 \begin{prop} The complex-form of the
 	$E_6^*$ lattice \cite{conway1984voronoi}  is equivalent to  $\mathcal{E}_{3}$.
 \end{prop} 
 \begin{IEEEproof}
 	Following \cite{conway1984voronoi}, the complex lattice basis of $E_6^*$ can be written as
 	\begin{align}
 	&\bar{\mathbf{B}}_{E_6^*}=\left[\begin{array}{ccc}
 	\sqrt{-3} & 1 &1\\
 	0 & -1 &0  \\
 	0 & 0 &-1 
 	\end{array}\right].
 	\end{align}
 	This lattice is generated by setting the sum of coordinates  as $(1 + \omega)\omega \mathbb{Z}[\omega]$. So we have $\mathcal{E}_{3} \approxeq \omega \mathcal{E}_{3} = E_6^*$.
 \end{IEEEproof}

\begin{prop} The $4$-dimensional checkerboard lattice $D_4$ satisfies
	$D_4 \approxeq \Psi(\mathcal{G}_{2})$, and the $8$-dimensional Gosset lattice $E_8$ satisfies $E_8 \approxeq \Psi(\mathcal{G}_{4,1+i}^+)$.
\end{prop}

\begin{IEEEproof}
	With reference to Eq. (\ref{eq:zizlRelation}), 
	the real-valued basis of the $\mathbb{Z}[i]$-based lattice is
		\begin{equation}
	\mathbf{B}_{\Psi(\bar{\Lambda})}=\left[\begin{array}{cc}
	\mathfrak{R}\left(\bar{\mathbf{B}}\right) & -\mathfrak{I}\left(\bar{\mathbf{B}}\right)\\
	\mathfrak{I}\left(\bar{\mathbf{B}}\right) & \mathfrak{R}\left(\bar{\mathbf{B}}\right)
	\end{array}\right].
	\end{equation}
	Then we have 
		\begin{align}
	\mathbf{B}_{\Psi(\mathcal{G}_{2})} &=\left[\begin{array}{cccc}
	1 & 0 &0 & 0 \\
	0 & 1 &1 & 1 \\
	0 & 0 &1 & 0 \\
	-1 & -1 &0 & 1 
	\end{array}\right] \\
		\mathbf{B}_{D_4} &=\left[\begin{array}{cccc}
	1 & 0 &0 & 0 \\
	0 & 1 &0 & 0 \\
	0 & 0 &1 & 0 \\
	1 & 1 &1 & 2 
	\end{array}\right].
	\end{align}
	Thus $\mathbf{B}_{\Psi(\mathcal{G}_{2})}$ equals to $\mathbf{B}_{D_4}$ up to reflection and unimodular multiplication.
	 
	Regarding $E_8$, its complex-valued lattice basis \cite{shum15} is:
	\begin{align}
	&\bar{\mathbf{B}}_{{E}_8}=\left[\begin{array}{cccc}
	1+i & 0 &0 & 0 \\
	1+i & -2 &0 & 0 \\
	1+i & 2i &-2 & 0 \\
	1+i & 0 &2i & 2+2i 
	\end{array}\right].
	\end{align}
	Then we have  $2\bar{\mathbf{B}}_{\mathcal{G}_{4,1+i}^+}\bar{\mathbf{U}}=\bar{\mathbf{B}}_{{E}_8}$, in which
	\begin{align}
	&\bar{\mathbf{B}}_{\mathcal{G}_{4,1+i}^+}=\left[\begin{array}{cccc}
	1/(1+i) & 0 &0 & 0 \\
	1/(1+i) & 1 &0 & 0 \\
	1/(1+i) & 0 &1 & 0 \\
	(1+4i)/(1+i) & i &i & 1+i
	\end{array}\right]\\
	&\bar{\mathbf{U}}=\left[\begin{array}{cccc}
	i & 0 &0 & 0 \\
	0 & -1 &0 & 0 \\
	0 & i &-1 & 0 \\
	-2i & 1 &1+i & 1 
	\end{array}\right].
	\end{align}
	Since $\bar{\mathbf{U}}$ is unimodular, the proposition is proved.
\end{IEEEproof}

\section{\label{sec:DecodeGoodComplexLattices}The Quantization Algorithms}
Implementing the proposed complex-lattice quantizers requires solving the associated CVP efficiently. This section presents algorithms for $Q_{\mathcal{E}_{m}}$, $Q_{\mathcal{G}_{m}}$, $Q_{\mathcal{E}_{m,2}^+}$, $Q_{\mathcal{E}_{m,1+\omega}^+}$, $Q_{\mathcal{G}_{m,2}^+}$, and $Q_{\mathcal{G}_{m,1+i}^+}$.
 In a high level, $Q_{\mathcal{E}_{m}}$ and $Q_{\mathcal{G}_{m}}$ both start from component-wise quantization, followed by modifying the coefficients to meet the lattice properties. $Q_{\mathcal{E}_{m,2}^+}$, $Q_{\mathcal{E}_{m,1+\omega}^+}$, $Q_{\mathcal{G}_{m,2}^+}$, and $Q_{\mathcal{G}_{m,1+i}^+}$ employ the structure of cosets.
 
%


\subsection{Algorithm of $Q_{\mathcal{E}_{m}}$}

  \begin{figure*}[t!]
	\centering
	\includegraphics[width=0.9\textwidth]{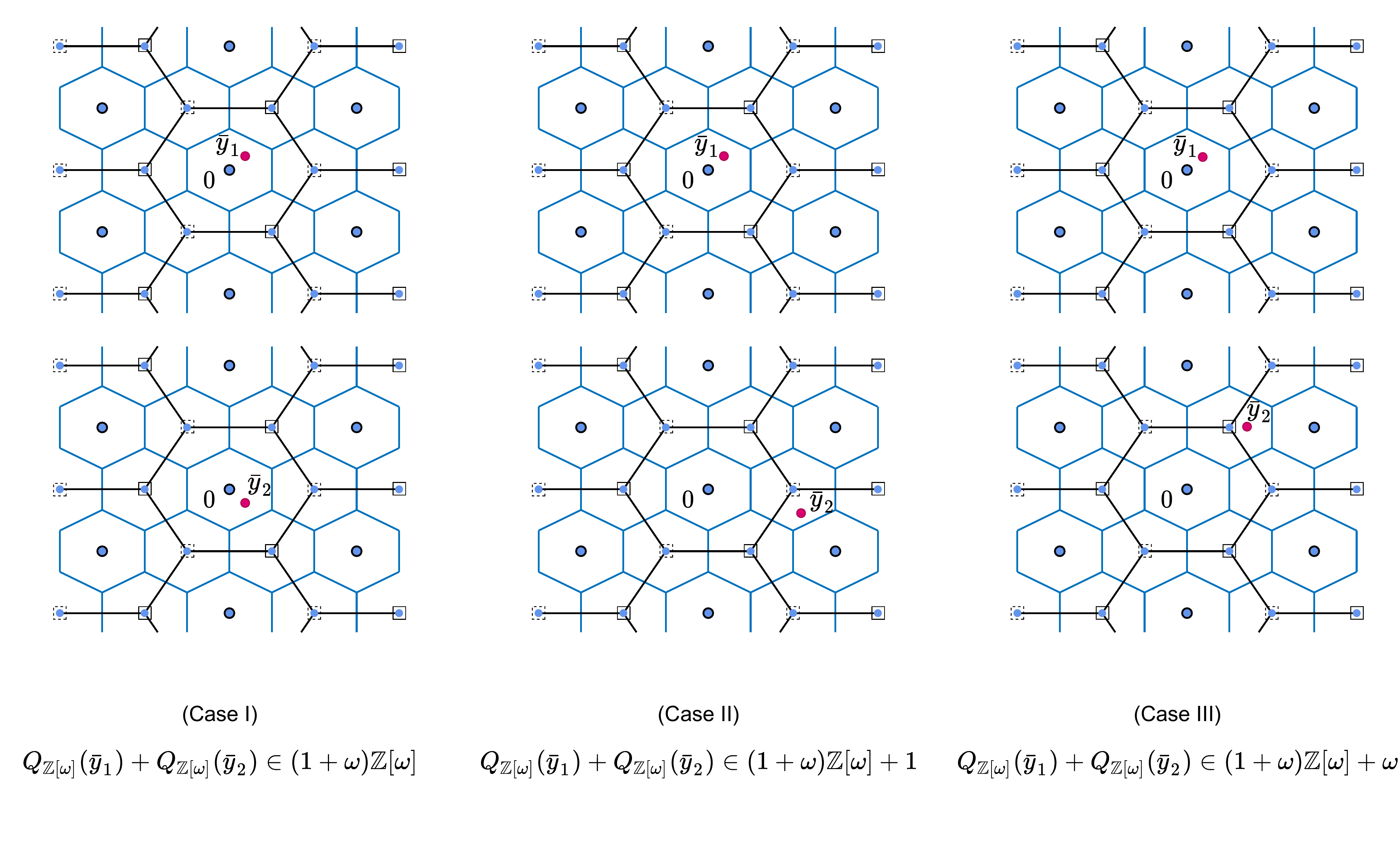}
	\caption{An illustrative example of quantizing different $\bar{\mathbf{y}}$ over  $\mathcal{E}_{2}$. Blue points denote $\mathbb{Z}[\omega]
		$, points enclosed with solid black circles denote $(1+\omega)\mathbb{Z}[\omega]$, points enclosed with    dash squares and solid squares respectively denote $(1+\omega)\mathbb{Z}[\omega]+1$ and $(1+\omega)\mathbb{Z}[\omega]+\omega$.}
	\label{figDwTheory}
\end{figure*}
 

By using element-wise quantization of $\bar{\mathbf{y}}$ over $\mathbb{Z}[\omega]$, we obtain
\begin{equation}
\bar{\mathbf{x}}=Q_{\mathbb{Z}[\omega]^m}(\bar{\mathbf{y}}).
\end{equation}
In case of a tie, choose the Eisenstein integer with the smallest absolute value.  
Then 
 we can add a perturbation vector $(\delta_1,\delta_2,...,\delta_m) \in \mathbb{Z}[\omega]^m$ to $\bar{\mathbf{x}}$, such that 
\begin{equation}
 \bar{\mathbf{x}}+ (\delta_1,\delta_2,...,\delta_m) \in \mathcal{E}_{m}, 
\end{equation}
while making $\left\|\bar{\mathbf{x}}+ (\delta_1,\delta_2,...,\delta_m) - \bar{\mathbf{y}} \right\|^2$ as small as possible. 
For each position $k$, obviously the smallest perturbation is  
$\delta_k=0$, and the second smallest is  $\delta_k \in \mathcal{S}_1 \cup \mathcal{S}_2$,
where
\begin{align}
\mathcal{S}_1 &=\{\omega^0,\, \,\omega^2,\,\omega^4\}=\{1,\, \,-1+\omega,\,-\omega\}, \\
\mathcal{S}_2 &=\{\omega,\,\omega^3,\,\omega^5\}=\{\omega,\,-1,\,-\omega+1\},
\end{align}
and $\mathcal{S}_1 \cup \mathcal{S}_2$ denotes the set of units in $\mathbb{Z}[\omega]$.
 The number of nonzero $\delta_k$, denoted as $\alpha$, can be $0, 1, 2, ..., m$. 

Due to the fact that 
\begin{equation}\label{eq_parZw03}
\mathbb{Z}[\omega]=(1+\omega)\mathbb{Z}[\omega] \cup \lbrace (1+\omega)\mathbb{Z}[\omega]+ 1 \rbrace \cup  \left\lbrace (1+\omega)\mathbb{Z}[\omega]+ \omega \right\rbrace,
\end{equation}
the summation of coefficients in $\bar{\mathbf{x}}$ consists of three cases:
\begin{align}
\sum_{k=1}^{m}\bar{x}_k &\in  (1+\omega)\mathbb{Z}[\omega], \,\, \text{(Case I)} \\
\sum_{k=1}^{m}\bar{x}_k &\in  (1+\omega)\mathbb{Z}[\omega]+1, \,\, \text{(Case II)}\\
\sum_{k=1}^{m}\bar{x}_k &\in  (1+\omega)\mathbb{Z}[\omega]+\omega. \,\, \text{(Case III)}
\end{align}
An example of $m=2$ is shown in Fig. \ref{figDwTheory}.
The algorithm proceeds according to the divided cases.
\begin{enumerate}
	\item    Case I: The perturbation vector should meet the requirement of $\sum_{k=1}^{m}\delta_k \in  (1+\omega)\mathbb{Z}[\omega]$.   Since $\bar{\mathbf{x}}$ is already the closest possible vector to $\bar{\mathbf{y}}$, we have $\alpha=0$ and
	the output vector is given by 
	$\hat{\mathbf{v}}=\bar{\mathbf{x}}$.
	\item Case II: The perturbation vector should meet the requirement of $\sum_{k=1}^{m}\delta_k \in  (1+\omega)\mathbb{Z}[\omega]+\omega$. Then we have $\alpha>0$.
	If $\alpha=1$, since
	\begin{align}
		&\omega, \omega^3, \omega^5 \in (1+\omega)\mathbb{Z}[\omega]+\omega\\
		&\omega^0, \omega^2, \omega^4 \notin (1+\omega)\mathbb{Z}[\omega]+\omega,
	\end{align}
	 we should choose one   $\delta_k \in \mathcal{S}_2$ to perturb $\bar{\mathbf{x}}$. 
To decide the value of $k$, we calculate the residue coefficient
	\begin{equation}\label{eqqs1}
	\bar{{r}}_k'=Q_{\mathcal{S}_2}(\bar{{y}}_k-\bar{x}_k) 
	\end{equation}
	and the incremental distance
	\begin{equation}
	L_k'=\left|\bar{x}_k + \bar{{r}}_k' - \bar{{y}}_k\right|^2 - \left|\bar{x}_k - \bar{{y}}_k\right|^2 
	\end{equation}
	for $k=1, ..., m$.
	The position with the smallest incremental distance is the one that we should change, as it leads to the closest vector. By defining $k^*= \arg \min_k L_k'$, the output candidate $\bar{\mathbf{x}}'$ is given by $\bar{x}_k'=\bar{x}_k$ for $k\neq k^*$, and $\bar{x}_k'=\bar{x}_k + \bar{{r}}_k'$ for $k=k^*$.
	
	If $\alpha=2$, due to the fact that
	\begin{align}
	&u+v \in (1+ \omega)\mathbb{Z}[\omega]+\omega \text{ if } u\in\mathcal{S}_1, v\in \mathcal{S}_1\\
	&u+v \notin (1+\omega)\mathbb{Z}[\omega]+\omega  \text{ if } u\in\mathcal{S}_1, v\in \mathcal{S}_2 \\
	&u+v \notin (1+\omega)\mathbb{Z}[\omega]+\omega  \text{ if } u\in\mathcal{S}_2, v\in \mathcal{S}_2,
	\end{align}
 we should choose two  $\delta_k$ from $\mathcal{S}_1$ to perturb $\bar{\mathbf{x}}$. 
Similarly, we calculate the residue coefficient
	\begin{equation}\label{eqqs2}
	\bar{{r}}_k''=Q_{\mathcal{S}_1}(\bar{{y}}_k-\bar{x}_k) 
	\end{equation}
	and the incremental distance
	\begin{equation}
L_k''=	\left|\bar{x}_k + \bar{{r}}_k'' - \bar{{y}}_k\right|^2 - \left|\bar{x}_k - \bar{{y}}_k\right|^2 
	\end{equation}
		for $k=1, ..., m$.
	Then we sort $\left\lbrace L_1'', ..., L_m''\right\rbrace$ in ascending order. Denote the two positions with the smallest incremental distance as $k_1^*$ and  $k_2^*$ respectively. The output candidate $\bar{\mathbf{x}}''$ is given by $\bar{x}_k''=\bar{x}_k$ for $k\neq k^*$, and $\bar{x}_k''=\bar{x}_k + \bar{{r}}_k''$ for $k=k_1^*$ and $k=k_2^*$.

	If $\alpha=3$, since
	\begin{align}
  &u+v+s \in (1+\omega)\mathbb{Z}[\omega]+\omega   \text{ if } u\in\mathcal{S}_1, v\in \mathcal{S}_2,  s\in \mathcal{S}_2 \\
	&u+v+s \notin (1+\omega)\mathbb{Z}[\omega]+\omega  \text{ if } u\in\mathcal{S}_1, v\in \mathcal{S}_1,  s\in \mathcal{S}_1 \\
	&u+v+s \notin (1+\omega)\mathbb{Z}[\omega]+\omega   \text{ if } u\in\mathcal{S}_2, v\in \mathcal{S}_2,  s\in \mathcal{S}_2 \\
	&u+v+s \notin (1+\omega)\mathbb{Z}[\omega]+\omega   \text{ if } u\in\mathcal{S}_1, v\in \mathcal{S}_2,  s\in \mathcal{S}_1,
	\end{align}
	the feasible perturbation vector $(\delta_1,\delta_2,...,\delta_m)$ contains two elements from $\mathcal{S}_2$ and one element from $\mathcal{S}_1$. Thus its corresponding perturbed candidate is no better than $\bar{\mathbf{x}}'$ that only uses $\mathcal{S}_2$ to perturb once.
	
	If $\alpha=4, 5, ..., m$, since $	\left|\bar{{x}}_{k^*}' + \bar{{r}}_{k^*} - \bar{{y}}_{k^*}\right|^2 - \left|\bar{{x}}_{k^*}' - \bar{{y}}_{k^*}\right|^2 \leq 1$, and $\min_k L_k' \geq 1/4$, $\min_k L_k'' \geq 1/4$, by perturbing $4$ or more positions of $\bar{\mathbf{x}}$, its distance to $\bar{\mathbf{y}}$ is no smaller than $\left\|\bar{\mathbf{x}}'- \bar{\mathbf{y}}\right\|^2$.
 
 Since $\alpha \leq m$, the $\bar{\mathbf{x}}''$ only exists when $m\geq 2$. 
	Summarizing Case II, when $m=1$, the algorithm outputs $\hat{\mathbf{v}}=\bar{\mathbf{x}}'$; when $m\geq 2$,
	 the algorithm outputs $\hat{\mathbf{v}}=\bar{\mathbf{x}}'$ if $\left\|\bar{\mathbf{x}}' -\bar{\mathbf{y}}\right\|^2 \leq \left\|\bar{\mathbf{x}}'' -\bar{\mathbf{y}}\right\|^2$, and $\hat{\mathbf{v}}=\bar{\mathbf{x}}''$ otherwise.
	\item Case III: The perturbation vector should meet the requirement of $\sum_{k=1}^{m}\delta_k \in  (1+\omega)\mathbb{Z}[\omega]+1$. The operations are similar to those in Case II, except that the roles of  $\mathcal{S}_1$ and $\mathcal{S}_2$ are swapped. The quantization functions become $Q_{\mathcal{S}_1}$ in Eq. (\ref{eqqs1}), and $Q_{\mathcal{S}_2}$ in Eq. (\ref{eqqs2}).
\end{enumerate}
 
The pseudocode of the  quantization algorithm is summarized in Algorithm \ref{AlgLQEM}. 
 \begin{algorithm}[t!]
 	\KwIn{A query vector $\bar{\mathbf{y}}$.}
 	\KwOut{The closest vector $\hat{\mathbf{v}}$ of $\bar{\mathbf{y}}$ in $\mathcal{E}_{m}$.}    
$\bar{\mathbf{x}}=Q_{\mathbb{Z}[\omega]^m}(\bar{\mathbf{y}})$\;
\If{$\sum_{k=1}^{m}\bar{x}_k \in  (1+\omega)\mathbb{Z}[\omega]$}{$\hat{\mathbf{v}}=\bar{\mathbf{x}}$}
\Else{
	\If{$\sum_{k=1}^{m}\bar{x}_k\in  (1+\omega)\mathbb{Z}[\omega]+1$}{$\mathcal{S}_1 =\{\omega^0,\, \,\omega^2,\,\omega^4\}$\;
	$\mathcal{S}_2 =\{\omega,\,\omega^3,\,\omega^5\}$}
	\Else{$\mathcal{S}_2 =\{\omega^0,\, \,\omega^2,\,\omega^4\}$\;
		$\mathcal{S}_1 =\{\omega,\,\omega^3,\,\omega^5\}$}
	\For{$k=1, ..., m$}{
$\bar{{r}}_k'=Q_{\mathcal{S}_2}(\bar{{y}}_k-\bar{x}_k)$	\;
$L_k'= 	\left|\bar{x}_k + \bar{{r}}_k' - \bar{{y}}_k\right|^2 - \left|\bar{x}_k - \bar{{y}}_k\right|^2$\;
}
$k^*= \arg \min_k L_k'$\;
$\bar{\mathbf{x}}'=\bar{\mathbf{x}}$\;
$\bar{{x}}_{k^*}'\leftarrow \bar{{x}}_{k^*}' + \bar{{r}}_{k^*}'$\;

\If{$m=1$}{	$\hat{\mathbf{v}}=\bar{\mathbf{x}}'$}\Else{\For{$k=1, ..., m$}{
		$\bar{{r}}_k''=Q_{\mathcal{S}_1}(\bar{{y}}_k-\bar{x}_k)$	\;
		$L_k''= 	\left|\bar{x}_k + \bar{{r}}_k'' - \bar{{y}}_k\right|^2 - \left|\bar{x}_k - \bar{{y}}_k\right|^2$\;
	}
	$k_{1}^*= \arg \min_k L_k''$;
	$k_{2}^*= \arg \min_k L_k''\backslash L_{k_{1}^*}''$\;
	$\bar{\mathbf{x}}''=\bar{\mathbf{x}}$\;
	$\bar{{x}}_{k_{1}^*}''\leftarrow \bar{{x}}_{k_{1}^*}'' + \bar{{r}}_{k_{1}^*}''$; 
	$\bar{{x}}_{k_{2}^*}''\leftarrow \bar{{x}}_{k_{2}^*}'' + \bar{{r}}_{k_{2}^*}''$\;
	\If{$\left\|\bar{\mathbf{x}}' -\bar{\mathbf{y}}\right\|^2 \leq \left\|\bar{\mathbf{x}}'' -\bar{\mathbf{y}}\right\|^2$}{
		$\hat{\mathbf{v}}=\bar{\mathbf{x}}'$
	}
	\Else{
		$\hat{\mathbf{v}}=\bar{\mathbf{x}}''$
}}
}
\caption{The closest vector algorithm of  $Q_{\mathcal{E}_{m}}$.}
 	\label{AlgLQEM}  
 \end{algorithm}

  
\subsection{Algorithm of $Q_{\mathcal{G}_{m}}$ \label{cardGiDecoding}}
To begin, we quantize  $\bar{\mathbf{y}}$ with respect to $\mathbb{Z}[i]^m$:
\begin{equation}
\bar{\mathbf{x}}=Q_{\mathbb{Z}[i]^m}(\bar{\mathbf{y}}).
\end{equation}
In case of a tie, choose the Gaussian integer with the smallest absolute value.  
Since
\begin{equation}\label{eq_parZi3}
\mathbb{Z}[i]=(1+i)\mathbb{Z}[i] \cup \lbrace (1+i)\mathbb{Z}[i]+ 1 \rbrace,
\end{equation}
the summation of coefficients in $\bar{\mathbf{x}}$ consists of two cases:
\begin{align}
\sum_{k=1}^{m}\bar{x}_k &\in  (1+i)\mathbb{Z}[i], \,\, \text{(Case I)} \\
\sum_{k=1}^{m}\bar{x}_k &\in  (1+i)\mathbb{Z}[i]+1, \,\, \text{(Case II)}
\end{align}
In Case I, $\bar{\mathbf{x}}$ is the closest vector of $\mathcal{G}_{m}$ to $\bar{\mathbf{y}}$. In Case II, one should perturb $\bar{\mathbf{x}}$ such that the perturbed vector $ \bar{\mathbf{x}}+ (\delta_1,\delta_2,...,\delta_m)\in \mathcal{G}_{m}$. For each component $\bar{x}_k$, the smallest perturbation is $\delta_k \in \{i,\,i^2,\,i^3,\,i^4\}$. As it suffices to perturb only one coefficient of $\bar{\mathbf{x}}$, the algorithm can be designed to search the perturbed position that
makes $\left\| \bar{\mathbf{x}}+  (\delta_1,\delta_2,...,\delta_m)-\bar{\mathbf{y}}\right\|^2$ as small as possible. 
The pseudocode of the  quantization algorithm of $Q_{\mathcal{G}_{m}}$ is summarized in Algorithm \ref{AlgLQGM}. 
 
\begin{algorithm}[t!]
	\KwIn{A query vector $\bar{\mathbf{y}}$.}
	\KwOut{The closest vector $\hat{\mathbf{v}}$ of $\bar{\mathbf{y}}$ in $\mathcal{G}_{m}$.}    
	$\bar{\mathbf{x}}=Q_{\mathbb{Z}[i]^m}(\bar{\mathbf{y}})$\;
	\If{$\sum_{k=1}^{m}\bar{x}_k \in  (1+i)\mathbb{Z}[i]$}{$\hat{\mathbf{v}}=\bar{\mathbf{x}}$}
	\Else{
		$\mathcal{S}=\{i,\,i^2,\,i^3,\,i^4\}$\;
		\For{$k=1, ..., m$}{
			$\bar{{r}}_k'=Q_{\mathcal{S}}(\bar{{y}}_k-\bar{x}_k)$	\;
			$L_k'= 	\left|\bar{x}_k + \bar{{r}}_k' - \bar{{y}}_k\right|^2 - \left|\bar{x}_k - \bar{{y}}_k\right|^2$\;
		}
	$k^*= \arg \min_k L_k'$\;
	$\bar{\mathbf{x}}'=\bar{\mathbf{x}}$\;
	$\bar{{x}}_{k^*}'\leftarrow \bar{{x}}_{k^*}' + \bar{{r}}_{k^*}'$\;
	$\hat{\mathbf{v}}=\bar{\mathbf{x}}'$
	}
	\caption{The closest vector algorithm of $Q_{\mathcal{G}_{m}}$.}
	\label{AlgLQGM}  
\end{algorithm}


\subsection{Algorithms of  $Q_{\mathcal{E}_{m,2}^+}$, $Q_{\mathcal{E}_{m,1+\omega}^+}$, $Q_{\mathcal{G}_{m,2}^+}$ and $Q_{\mathcal{G}_{m,1+i}^+}$}

If a lattice is built from the union of cosets, i.e.,  
\begin{equation}
	\Lambda= \cup_{t} (\mathbf{g}_t+\Lambda'),
\end{equation}
then  $Q_{\Lambda'}$ can be used as the basis of $Q_{\Lambda}$. To be concise, we have
%
%
\begin{align}\label{union_decode1}
Q_{\Lambda }(\mathbf{y})  =  Q_{\Lambda'+\mathbf{g}_{t^*}}(\mathbf{y}), \,
t^*=\mathrm{argmin}_{t} \left\| \mathbf{y} -	Q_{\Lambda'+\mathbf{g}_t}(\mathbf{y}) \right\|. 
\end{align}

By representing  $\mathcal{E}_{m,2}^+$ and $\mathcal{E}_{m,1+\omega}^+$ as unions of $\mathcal{E}_{m}$-cosets,  
 $\mathcal{G}_{m,2}^+$ and $\mathcal{G}_{m,1+i}^+$ as unions of $\mathcal{G}_{m}$-cosets,  the quantization algorithms of $Q_{\mathcal{E}_{m,2}^+}$, $Q_{\mathcal{E}_{m,1+\omega}^+}$, $Q_{\mathcal{G}_{m,2}^+}$, $Q_{\mathcal{G}_{m,1+i}^+}$  follow from Eq. (\ref{union_decode1}).

 
 \subsection{Computational Complexity}
 The overall complexity of a quantization algorithm over $\Lambda\in \mathbb{R}^n$ can be given as
 \begin{align}
 	\mathrm{Comp}(Q_{\Lambda}) = \sum_{k=1}^{S} f_n(k), 
 \end{align}
 where $S$ represents the number of visited lattice vectors, and $f_n(k)$ denotes the number of elementary operations (referred to as flops, including additions, subtractions,  multiplications and scalar quantization) that the algorithm performs in the $k$th visited vector.
 
 Regarding the Algorithm 1 for $Q_{\mathcal{E}_{m}}$, if it terminates in Step 3, then $S=1$; otherwise $S=3$. In the worst case of $S=3$, we have 
 \begin{align}
 	 	\mathrm{Comp}(Q_{\mathcal{E}_{m}}) &= f_{2m}(1) + f_{2m}(2) +f_{2m}(3)  \\
 	 	&\approx 2m + m\times(3+6)+  m\times(3+6) \\
 	 	&= 20m,
 \end{align}
 where $f_{2m}(1)$ is from Step 1, $f_{2m}(2)$ is from Steps 11 to 13, and $f_{2m}(3)$ is from Steps 20 to 22.
 
 Since $|\mathcal{E}_{m,1+\omega}^+/\mathcal{E}_{m}|=3$, $|\mathcal{E}_{m,2}^+/\mathcal{E}_{m}|=4$, we have 	 	$\mathrm{Comp}(Q_{\mathcal{E}_{m,1+\omega}^+})=3 \times	\mathrm{Comp}(Q_{\mathcal{E}_{m}}) \approx 60m$, $\mathrm{Comp}(Q_{\mathcal{E}_{m,2}^+})=4 \times	\mathrm{Comp}(Q_{\mathcal{E}_{m,2}^+}) \approx 80m$. In the same vein, 
regarding Algorithm 2 for $Q_{\mathcal{G}_{m}}$, $S$ can be at most $2$, and $\mathrm{Comp}(Q_{\mathcal{G}_{m}}) \approx 11m$. 
 As $|\mathcal{G}_{m,1+i}^+/\mathcal{G}_{m}|=2$, $|\mathcal{G}_{m,2}^+/\mathcal{G}_{m}|=4$, we have 	 	$\mathrm{Comp}(Q_{\mathcal{G}_{m,1+i}^+}) \approx 22m$, $\mathrm{Comp}(Q_{\mathcal{G}_{m,2}^+}) \approx 44m$. 
 
 
The proposed algorithms have an extraordinary feature: 
 the number of visited lattice vectors $S$ is independent of the lattice dimension $m$. Specifically, we have $3$, $9$ and $12$ visited vectors for  $\mathcal{E}_{m}$, $\mathcal{E}_{m,1+\omega}^+$ and  $\mathcal{E}_{m,2}^+$;  $2$, $4$ and $8$ visited vectors for  $\mathcal{G}_{m}$, $\mathcal{G}_{m,1+i}^+$ and  $\mathcal{G}_{m,2}^+$.
 This feature saves a large amount of computational complexity over other possible alternatives. E.g., the universal enumeration algorithm (cf. \cite{Micciancio2002,Hassibi2005,scn/AlbrechtCDDPPVW18}) has an exponential number of $S$, while the quantization algorithm  in  \cite{conway1984voronoi} (i.e., to factorize $\mathcal{E}_{m}$ as the union of $(1+\omega)\mathbb{Z}[\omega]^m$-cosets) involves $S=3^{m-1}$ visited vectors. 
 
 \section{Numerical Evaluation}
 Theoretically analyzing the exact NSM is complicated as it requires a complete description of the Voronoi regions of the generalized checkerboard lattices. With the aid of the proposed fast quantization algorithms, the Monte Carlo integration method   \cite{conway1984voronoi} can be employed to  
 compute the NSM with high accuracy. To foster reproducible research, our programs used in
 the simulations are of open source and freely available at GitHub\footnote{https://github.com/shx-lyu/LatticeQuantizer}.
 
\subsection{Method}
To begin, we review the Monte Carlo integration method \cite{conway1984voronoi} for real-valued lattices.
Let $\mathbf{b}_1, ..., \mathbf{b}_{2m}$ be linearly independent vectors spanning the lattice $\Lambda$, and let $u_1, ..., u_{2m}$ be independent random numbers, uniformly distributed between $0$ and $1$. Then $\mathbf{y}=\sum_{k=1}^{2m}\mathbf{b}_k u_k$ is   uniformly distributed over the fundamental parallelepiped generated by $\mathbf{b}_1, ..., \mathbf{b}_{2m}$. Then $\mathbf{y}-Q_\Lambda(\mathbf{y})$ is uniformly distributed over the Voronoi region $\mathcal{V}_{\Lambda}$.

 With $T=gh$ random points $\mathbf{y}^{(0)}, ..., \mathbf{y}^{(T-1)}$ selected in the manner just described,   the estimated NSM is given by $\hat{G}_{n}(\Lambda)= \frac{\hat{I}}{n \rm{Vol}({\Lambda})^{1+\frac{2}{n}}}$, where
\begin{equation}
	\hat{I}=\frac{1}{T} \sum_{t=1}^{T} \left\|\mathbf{y}^{(t)} - Q_\Lambda(\mathbf{y}^{(t)})\right\|^2
\end{equation}
  is an estimate of $I={\int_{\mathbf{x}\in \mathcal{V}_{\Lambda} } ||\mathbf{x}||^2 \mathrm{d}\mathbf{x}}$. 
  
 $\mathbf{y}^{(0)}, ..., \mathbf{y}^{(T-1)}$ are further partitioned into $g$ sets, each has $h$ elements. Based on the jackknife estimator (see \cite{conway1984voronoi,sawyer2005resampling}), the standard deviation of $\hat{G}_{n}(\Lambda)$ is
  \begin{align}
  \hat{\sigma} &=\frac{1}{n \rm{Vol}({\Lambda})^{1+\frac{2}{n}}} \times  \nonumber \\
  & \sqrt{\frac{1}{g(g-1)} \sum_{s=0}^{g-1}\left(\frac{1}{h}\sum_{t=sh}^{(s+1)h-1} \left\|\mathbf{y}^{(t)} - Q_\Lambda(\mathbf{y}^{(t)})\right\|^2 - \hat{I}\right)^2}.
  \end{align}
Following the nomenclature in  \cite{conway1984voronoi,agrell98}, the confidence interval of the estimation is given by $\hat{G}_{n}(\Lambda)\pm 2   \hat{\sigma}$.
 
  With the complex-to-real transform $\Psi$ in Eq. (\ref{eq_rct}), one may employ a universal real-valued quantization algorithm $Q_{\Lambda}$ (e.g., enumeration) for the constructed lattices (e.g., $\Lambda=\Psi(\mathcal{E}_{m})$, $\Lambda=\Psi(\mathcal{E}_{m,2}^+)$), but the computational complexity is too high as it fails to utilize the algebraic structure of the generalized checkerboard lattices. Fortunately, the proposed  quantization algorithms for the complex lattices help to solve this issue. Specifically, let $\mathbf{y}^{(t)}=\Psi(\bar{\mathbf{y}}^{(t)})$, then we have
  \begin{align}
  	\left\|\mathbf{y}^{(t)} - Q_{{\Lambda}}(\mathbf{y}^{(t)})\right\|^2 &= \left\|\Psi^{-1}(\mathbf{y}^{(t)}) - Q_{\Psi^{-1}(\Lambda)}(\Psi^{-1}(\mathbf{y}^{(t)}))\right\|^2 \\
  	&=  	\left\|\bar{\mathbf{y}}^{(t)} - Q_{\bar{\Lambda}}(\bar{\mathbf{y}}^{(t)})\right\|^2
  \end{align}
  where $\bar{\mathbf{y}}^{(t)}= \sum_{k=1}^{m}\bar{\mathbf{b}}_k (u_k + \omega u_{k+m})$ and $\left[\bar{\mathbf{b}}_1, ..., \bar{\mathbf{b}}_m\right] $ denotes a generator matrix of $\bar{\Lambda}$.
  
  \begin{rem}
  	The proposed quantizers only correspond to even dimensional real-valued lattices. For odd dimensions, we can leverage Agrell and Allen's recent result \cite[Cor. 5]{agrell2022} about product lattices. 
  	For given lattices $\Lambda \in \mathbb{R}^{2m}$ and $\mathbb{Z} \in \mathbb{R}$, the  
  	product lattice $\Lambda_{\mathrm{opt}} = \Lambda \otimes a \mathbb{Z}$ with $a=\sqrt{12G_{2m}(\Lambda)} \mathrm{Vol}(\Lambda)^{\frac{1}{2m}}$ satisfies
  	\begin{align}
  	G_{2m+1}(\Lambda_{\mathrm{opt}})
  	&=G_{2m}(\Lambda)^{\frac{2m}{2m+1}} 12^{-\frac{1}{2m+1}} \\
  	\mathrm{Vol}(\Lambda_{\mathrm{opt}}) &=a \mathrm{Vol}(\Lambda).
  	\end{align}
  \end{rem}
  
\subsection{Performance}
 In the sequel, we set $T=5\times 10^7$ ($g=50, h=10^6$) in the Monte Carlo integration. The standard variance of each estimation satisfies $\hat{\sigma}\leq 1.65\times 10^{-6}$. The numerically evaluated NSMs keep $5$ decimal places, and those with theoretical exact values keep $9$ decimal places.
 
\begin{figure}[t!]
 	\centering
 	\subfigure[Eisenstein-integer based lattices.]{\includegraphics[width=.45\textwidth]{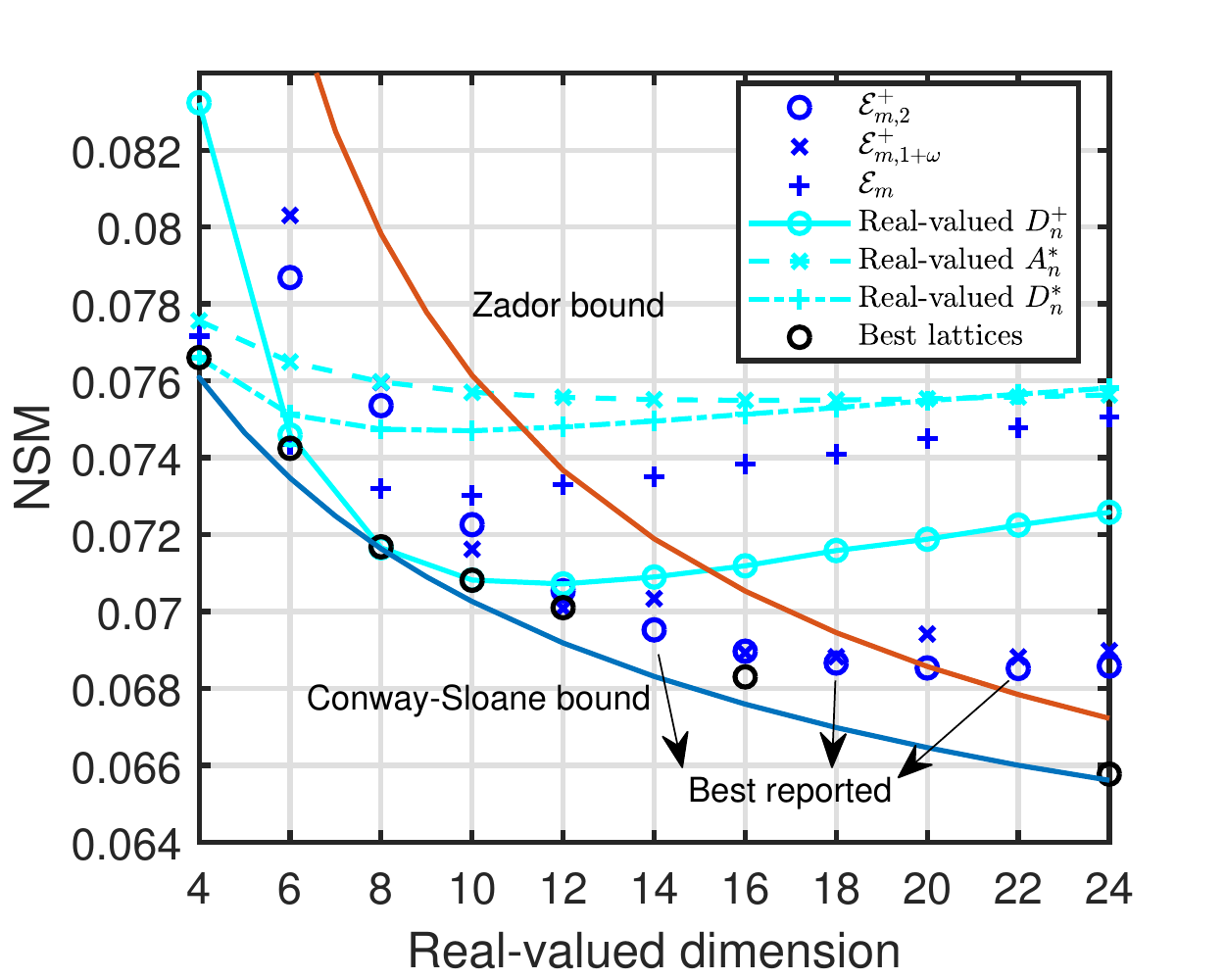}} 
 	\subfigure[Gaussian-integer based lattices.]{\includegraphics[width=.45\textwidth]{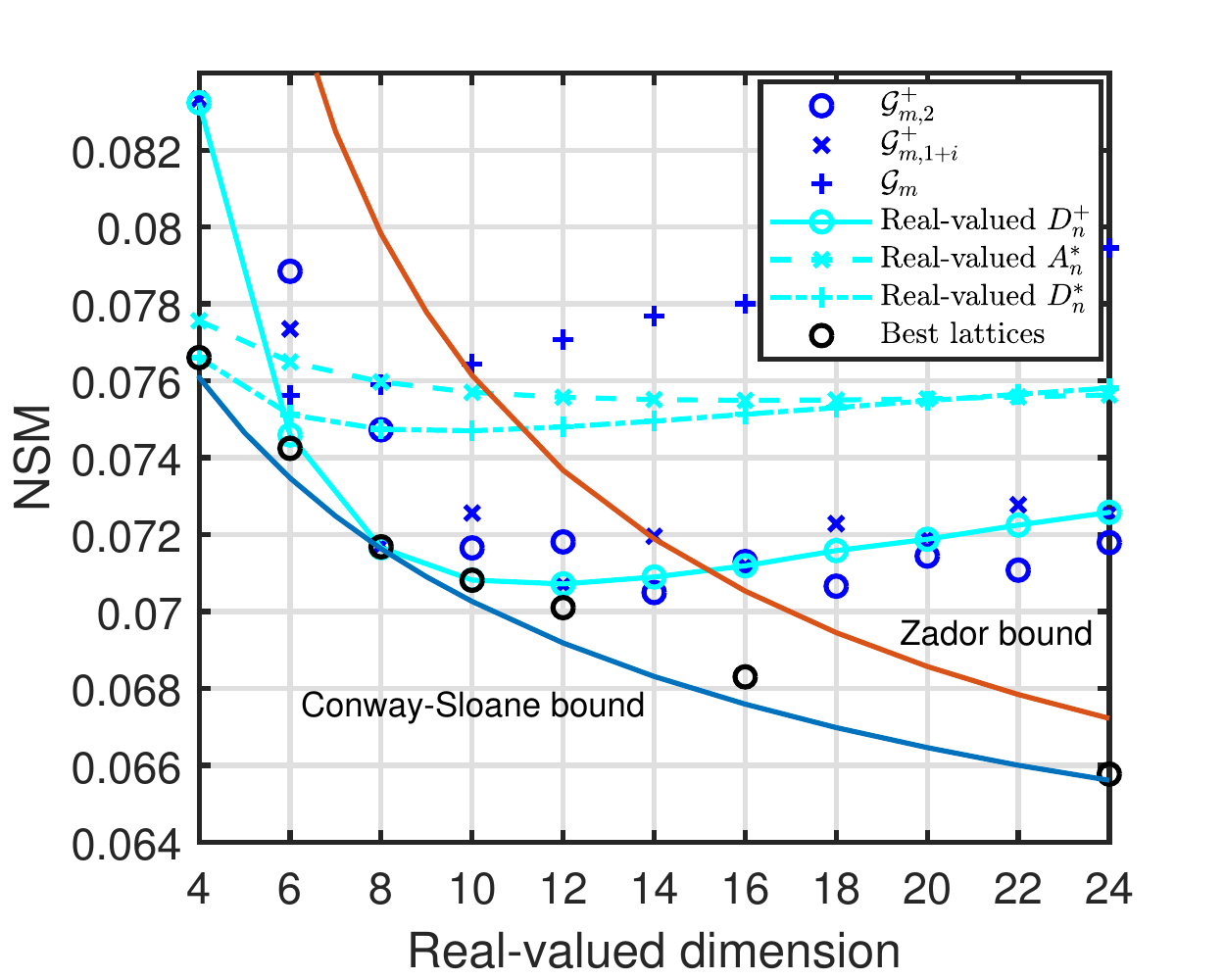}}
 	\caption{The NSM performance of the generalized checkerboard lattices.} 
 	\label{figDwiMoments}
 \end{figure}

 Fig. \ref{figDwiMoments} compares the NSM performance of the generalized checkerboard lattices with existing results. Benchmarks include the conjectured lower bound from Conway and Sloane \cite{tit/ConwayS85}, Zador's upper bound \cite{tit/Zador82}, root lattices $D_n^+$, $A_n^*$, $D_n^*$, and some typical best known lattices $D_4$, $D_8^+$, $K_{12}$, $\Lambda_{16}$, and $\Lambda_{24}$. 
 \begin{itemize}
 	\item  In Fig. \ref{figDwiMoments}-(a), it is shown that  $G_{14}(\Psi(\mathcal{E}_{7,2}^+))=0.06952$, $G_{18}(\Psi(\mathcal{E}_{9,2}^+))=0.06866$,  and
 	$G_{22}(\Psi(\mathcal{E}_{11,2}^+))=0.06853$, which attain the smallest reported NSMs in dimensions $14$, $18$, and $22$.
 	The product lattices \cite{agrell2022} based on $\Psi(\mathcal{E}_{7,2}^+)$, $\Psi(\mathcal{E}_{9,2}^+)$,  and $\Psi(\mathcal{E}_{11,2}^+)$ also yield the best reported NSMs, which are $0.07037$, $0.06936$,  and $0.06912$ in dimensions $15$, $19$, and $23$.
 	 In dimensions $n\leq 20$, the $\mathcal{E}_{m,2}^+$ quantizers lie below both the upper bound given by Zador \cite{tit/Zador82} and the quantizers based on $D_n^*$ and $A_n^*$.
%
 	Moreover, $\mathcal{E}_{m,2}^+$ outperforms $\mathcal{E}_{m,1+\omega}^+$ when the real-valued dimension is larger than $12$.
 	For comparison,
 	the NSM curve of $D_n^+$ reflects the performance limits of real-valued checkerboard-lattice cosets. 
 	\item  Fig. \ref{figDwiMoments}-(b) reveals the performance of  $\mathcal{G}_{m,2}^+$,  $\mathcal{G}_{m,1+i}^+$ and  $\mathcal{G}_{m}$ with the same benchmarks. These Gaussian integers-based checkerboard lattices fail to exhibit better NSMs than those based on Eisenstein integers except when $n=4$ and $n=8$.
 \end{itemize}

 Table \ref{tab_GcDimN2} summarizes a complete list of the best reported quantizers in dimensions $n\leq 24$.
 It is noteworthy that when $n> 24$, the proposed lattices cannot exhibit decreasing NSMs. The reason is that both the real-valued and complex-valued checkerboard lattices can be regarded as generating from the single-parity-check codes, in which the complex-valued extensions assist to make 
 approximately twice as large the best reported dimensions.
 
 
  
 \begin{table*}[th]
 	\centering
 	\caption{The best reported lattice quantizers.}
 	\label{tab_GcDimN2}
\begin{tabular}{c|c|c|c|c|c|c|c}
	\hline 
	\multicolumn{1}{c|}{Dimension} & \multicolumn{2}{c|}{Best previously reported} & \multicolumn{2}{c|}{Generic bounds} & \multicolumn{2}{c|}{Proposed} & \multicolumn{1}{c}{Better than}\tabularnewline
	\cline{2-7} 
$n$ & NSM & Lattice & Lower \cite{tit/ConwayS85} & Upper \cite{tit/Zador82}  & NSM & Lattice & Reported  \tabularnewline
	\hline 
	\hline 
	$1$ &  {$0.083333333$} & $\text{\ensuremath{\mathbb{Z}}}$ & $0.083333333$ & $0.500000000$ & $ $ & $ $ &   \tabularnewline
	$2$ & $0.080187537$ & $A_{2}$ & $0.080187537$ & $0.159154943$ & $0.080187537$ & $\Psi(\mathcal{{E}}_{1})$ & $=$  \tabularnewline
	$3$ & $0.078543281$ & $A_{3}^{*}$ & $0.077874985$ & $0.115802581$ & $0.081222715$ & $\Psi(\mathcal{{E}}_{1})\otimes a\mathbb{Z}$ &   \tabularnewline
	$4$ & $0.076603235$ & $D_{4}$ & $0.076087080$ & $0.099735570$ & $0.076603235$ & $\Psi(\mathcal{G}_{2})$ & $=$  \tabularnewline
	$5$ & $0.075625443$ & $D_{5}^{*}$ & $0.074654327$ & $0.091319469$ & $0.077904301$ & $\Psi(\mathcal{{G}}_{2}) \otimes a\mathbb{Z}$ &   \tabularnewline
	$6$ & $0.074243697$ & $E_{6}^{*}$ & $0.073474906$ & $0.086084334$ & $0.07424$ & $\Psi(\mathcal{{E}}_{3})$ & $=$  \tabularnewline
	$7$ & $0.073116493$ & $E_{7}^{*}$ & $0.072483503$ & $0.082478806$ & $0.07548$ & $\Psi(\mathcal{E}_{3})\otimes a\mathbb{Z}$ &   \tabularnewline
	$8$ & $0.071682099$ & $E_{8}$ & $0.071636064$ & $0.079824101$ & $0.071682099$ & $\Psi({\mathcal{G}}_{4,1+i}^{+})$ & $=$  \tabularnewline
	\hline 
	$9$ & $0.071622594$ & $AE_{9}$ & $0.070901661$ & $0.077775626$ & $0.072891732$ & $\Psi({\mathcal{G}}_{4,1+i}^{+}) \otimes a\mathbb{Z}$ &   \tabularnewline
	$10$ & $0.070813818$ & $D_{10}^{+}$ & $0.070257874$ & $0.076139300$ & $0.07162$ & $\Psi(\mathcal{E}_{5,1+\omega}^{+})$ &   \tabularnewline
	$11$ & $0.070426259$ & $A_{11}^{3}$ & $0.069688002$ & $0.074797093$ & $0.07261$ & $\Psi(\mathcal{E}_{5,1+\omega}^{+}) \otimes a\mathbb{Z}$ &   \tabularnewline
	$12$ & $0.070095600$ & $K_{12}$ & $0.069179323$ & $0.073672867$ & $0.07009$ & $\Psi(\mathcal{E}_{6,1+\omega}^{+})$ &    \tabularnewline
	$13$ & $0.071034583$ & $K_12 \otimes a\mathbb{Z}$ \cite{agrell2022} & $0.068721956$ & $0.072715163$ & $0.07103$ & $\Psi(\mathcal{E}_{6,1+\omega}^{+}) \otimes a\mathbb{Z}$ &   \tabularnewline
	$14$ & $0.071455542$ & $K_{12}\otimes aA_2$ \cite{agrell2022} & $0.068308096$ & $0.071887858$ & $\mathbf{{0.06952}}$ &  {${\Psi(\mathcal{E}_{7,2}^+)}$} & {Yes}  \tabularnewline
	$15$ & $0.071709124$ & $K_{12}\otimes aA_3^*$ \cite{agrell2022} & $0.067931488$ & $0.071164794$ & {$\mathbf{0.07037}$} &  {${\Psi(\mathcal{E}_{7,2}^+)\otimes a\mathbb{Z}}$} & {Yes}  \tabularnewline
	$16$ & $0.06830$ & $\Lambda_{16}$ & $0.067587055$ & $0.070526523$ & $0.06895$ & $\Psi(\mathcal{E}_{8}^{+})$ &   \tabularnewline
	\hline 
	$17$ & $0.06910$ & $\Lambda_{16}\otimes a\mathbb{Z}$ \cite{agrell2022} & $0.067270625$ & $0.069958259$ & $0.06972$ & $\Psi(\mathcal{E}_{8}^{+})\otimes a\mathbb{Z}$ &   \tabularnewline
	$18$ & $0.06953$ & $\Lambda_{16}\otimes aA_{2}$ \cite{agrell2022} & $0.066978741$ & $0.069448546$ & {$\mathbf{0.06866}$} &  {${\Psi(\mathcal{E}_{9,2}^+)}$} & {Yes}  \tabularnewline
	$19$ & $0.06982$ & $\Lambda_{16}\otimes aA_{3}^{*}$ \cite{agrell2022} & $0.066708503$ & $0.068988355$ & {$\mathbf{0.06936}$} &   {${\Psi(\mathcal{E}_{9,2}^+)\otimes a\mathbb{Z}}$} & {Yes}  \tabularnewline
	$20$ & $0.06769$ &  $(32,31)$ \cite{isit/KudryashovY10} & $0.066457468$ & $0.068570467$ &  {$0.06854$} & {$\Psi(\mathcal{E}_{10,2}^+)$} &   \tabularnewline
	$21$ & $0.06836$ & $ (32,31)\otimes a \mathbb{Z}$ \cite{agrell2022,isit/KudryashovY10} & $0.066223553$ & $0.068189035$ & {$0.06918$} &  {$\Psi(\mathcal{E}_{10,2}^+)\otimes a\mathbb{Z}$} &   \tabularnewline
	$22$ & $0.06987$ & $\Lambda_{16}\otimes aE_{6}^{*}$ \cite{agrell2022} & $0.066004976$ & $0.067839266$ &  {$\mathbf{0.06853}$} &  {${\Psi(\mathcal{E}_{11,2}^+)}$} & {Yes}  \tabularnewline
	$23$ & $0.06973$ & $\Lambda_{16}\otimes aE_{7}^{*}$ \cite{agrell2022} & $0.065800200$ & $0.067517194$ & {$\mathbf{0.06912}$} &  {${\Psi(\mathcal{E}_{11,2}^+)\otimes a\mathbb{Z}}$} & {Yes} \tabularnewline
	$24$ & $0.06577$ & $\Lambda_{24}$ & $0.065607893$ & $0.067219503$ & $0.06858$ & $\Psi(\mathcal{E}_{12,2}^+)$ &    \tabularnewline
	\hline 
\end{tabular}
 \end{table*}

\section{Conclusions}
Unlike Conway and Sloane's approach of lifting linear codes to complex lattice by complex Construction A \cite[Page 197]{conway1984voronoi}, the proposed $\mathcal{E}_{m}$ and $\mathcal{G}_{m}$ are constructed by algebraic equations, and this property is leveraged to design fast quantization algorithms. The best reported NSMs in dimensions $14$, $15$, $18$, $19$, $22$ and $23$ are due to $\mathcal{E}_{m,2}^+$, which is obtained by mimicking $D_n^+$. 

Since Eisenstein integers and Gaussian integers are the best two types of rings of imaginary quadratic fields, we believe that the proposed lattices based on these two rings already capture the highest potential when generalizing checkerboard lattices in quadratic fields.
The future work may study the generalization with quaternions.

\bibliographystyle{IEEEtranMine}
\bibliography{lib}

\begin{thebibliography}{10}
\providecommand{\url}[1]{#1}
\csname url@samestyle\endcsname
\providecommand{\newblock}{\relax}
\providecommand{\bibinfo}[2]{#2}
\providecommand{\BIBentrySTDinterwordspacing}{\spaceskip=0pt\relax}
\providecommand{\BIBentryALTinterwordstretchfactor}{4}
\providecommand{\BIBentryALTinterwordspacing}{\spaceskip=\fontdimen2\font plus
\BIBentryALTinterwordstretchfactor\fontdimen3\font minus
  \fontdimen4\font\relax}
\providecommand{\BIBforeignlanguage}[2]{{%
\expandafter\ifx\csname l@#1\endcsname\relax
\typeout{** WARNING: IEEEtran.bst: No hyphenation pattern has been}%
\typeout{** loaded for the language `#1'. Using the pattern for}%
\typeout{** the default language instead.}%
\else
\language=\csname l@#1\endcsname
\fi
#2}}
\providecommand{\BIBdecl}{\relax}
\BIBdecl

\bibitem{tit/Forney88a}
\BIBentryALTinterwordspacing
{G. David Forney Jr.}, ``Coset codes-{II}: Binary lattices and related codes,''
  \emph{{IEEE} Trans. Inf. Theory}, vol.~34, no.~5, pp. 1152--1187, 1988.
\BIBentrySTDinterwordspacing

\bibitem{Erez2004}
U.~Erez and R.~Zamir, ``Achieving 1/2 log {(1+SNR)} on the {AWGN} channel with
  lattice encoding and decoding,'' \emph{{IEEE} Trans. Inf. Theory}, vol.~50,
  no.~10, pp. 2293--2314, Oct. 2004.

\bibitem{tit/CampelloLB18}
\BIBentryALTinterwordspacing
A.~Campello, C.~Ling, and J.~Belfiore, ``Universal lattice codes for {MIMO}
  channels,'' \emph{{IEEE} Trans. Inf. Theory}, vol.~64, no.~12, pp.
  7847--7865, 2018.
\BIBentrySTDinterwordspacing

\bibitem{zamir2014lattice}
R.~Zamir, \emph{Lattice Coding for Signals and Networks: A Structured Coding
  Approach to Quantization, Modulation, and Multiuser Information
  Theory}.\hskip 1em plus 0.5em minus 0.4em\relax Cambridge University Press,
  2014.

\bibitem{DBLP:journals/tsp/ChoiNL18}
\BIBentryALTinterwordspacing
J.~Choi, Y.~Nam, and N.~Lee, ``Spatial lattice modulation for {MIMO} systems,''
  \emph{{IEEE} Trans. Signal Process.}, vol.~66, no.~12, pp. 3185--3198, 2018.
\BIBentrySTDinterwordspacing

\bibitem{lin2021lattice}
J.~Lin, J.~Qin, S.~Lyu, B.~Feng, and J.~Wang, ``Lattice-based
  minimum-distortion data hiding,'' \emph{IEEE Communications Letters}, 2021.

\bibitem{Conway1999}
J.~H. Conway and N.~J.~A. Sloane, \emph{Sphere Packings, Lattices and Groups},
  3rd~ed.\hskip 1em plus 0.5em minus 0.4em\relax Springer New York, 1999.

\bibitem{hales2005proof}
T.~C. Hales, ``A proof of the {Kepler} conjecture,'' \emph{Annals of
  mathematics}, vol. 162, no.~3, pp. 1065--1185, 2005.

\bibitem{viazovska2017sphere}
M.~S. Viazovska, ``The sphere packing problem in dimension 8,'' \emph{Annals of
  Mathematics}, pp. 991--1015, 2017.

\bibitem{cohn2017sphere}
H.~Cohn, A.~Kumar, S.~Miller, D.~Radchenko, and M.~Viazovska, ``The sphere
  packing problem in dimension $24$,'' \emph{Annals of Mathematics}, vol. 185,
  no.~3, pp. 1017--1033, 2017.

\bibitem{allen2021optimal}
B.~Allen and E.~Agrell, ``The optimal lattice quantizer in nine dimensions,''
  \emph{Annalen der Physik}, vol. 533, no.~12, p. 2100259, 2021.

\bibitem{isit/KudryashovY10}
\BIBentryALTinterwordspacing
B.~D. Kudryashov and K.~V. Yurkov, ``Near-optimum low-complexity lattice
  quantization,'' in \emph{{IEEE} International Symposium on Information
  Theory, {ISIT} 2010, June 13-18, 2010, Austin, Texas, USA}, pp. 1032--1036,
  2010.
\BIBentrySTDinterwordspacing

\bibitem{agrell2022}
E.~Agrell and B.~Allen, ``On the best lattice quantizers,'' \emph{arXiv
  preprint, arXiv:2202.09605}, 2022.

\bibitem{agrell98}
\BIBentryALTinterwordspacing
E.~Agrell and T.~Eriksson, ``Optimization of lattices for quantization,''
  \emph{{IEEE} Trans. Inf. Theory}, vol.~44, no.~5, pp. 1814--1828, 1998.
\BIBentrySTDinterwordspacing

\bibitem{lyu20im}
\BIBentryALTinterwordspacing
S.~Lyu, C.~Porter, and C.~Ling, ``Lattice reduction over imaginary quadratic
  fields,'' \emph{{IEEE} Trans. Signal Process.}, vol.~68, pp. 6380--6393,
  2020.
\BIBentrySTDinterwordspacing

\bibitem{Gan2009}
Y.~H. Gan, C.~Ling, and W.~H. Mow, ``Complex lattice reduction algorithm for
  low-complexity full-diversity {MIMO} detection,'' \emph{{IEEE} Trans. Signal
  Process.}, vol.~57, no.~7, pp. 2701--2710, 2009.

\bibitem{Micciancio2002}
D.~Micciancio and S.~Goldwasser, \emph{Complexity of Lattice Problems}.\hskip
  1em plus 0.5em minus 0.4em\relax Boston, MA: Springer, 2002.

\bibitem{shum15}
\BIBentryALTinterwordspacing
K.~W. Shum and Q.~T. Sun, ``Lattice network codes over optimal lattices in low
  dimensions,'' in \emph{Seventh International Workshop on Signal Design and
  its Applications in Communications, {IWSDA} 2015, Bengaluru, India}, pp.
  113--117, 2015.
\BIBentrySTDinterwordspacing

\bibitem{Sun2013}
Q.~T. Sun, J.~Yuan, T.~Huang, and K.~W. Shum, ``Lattice network codes based on
  {E}isenstein integers,'' \emph{{IEEE} Trans. Commun.}, vol.~61, no.~7, pp.
  2713--2725, Jul. 2013.

\bibitem{jerry2018}
\BIBentryALTinterwordspacing
Y.~Huang, K.~R. Narayanan, and P.~Wang, ``Lattices over algebraic integers with
  an application to compute-and-forward,'' \emph{{IEEE} Trans. Inf. Theory},
  vol.~64, no.~10, pp. 6863--6877, 2018.
\BIBentrySTDinterwordspacing

\bibitem{martinet2013perfect}
J.~Martinet, \emph{Perfect lattices in Euclidean spaces}.\hskip 1em plus 0.5em
  minus 0.4em\relax Springer Science \& Business Media, 2013, vol. 327.

\bibitem{conway1984voronoi}
J.~H. Conway and N.~J. Sloane, ``On the {V}oronoi regions of certain
  lattices,'' \emph{SIAM Journal on Algebraic Discrete Methods}, vol.~5, no.~3,
  pp. 294--305, 1984.

\bibitem{Hassibi2005}
B.~Hassibi and H.~Vikalo, ``On the sphere-decoding algorithm {I}. expected
  complexity,'' \emph{{IEEE} Trans. Signal Process.}, vol.~53, no. 8-1, pp.
  2806--2818, 2005.

\bibitem{scn/AlbrechtCDDPPVW18}
\BIBentryALTinterwordspacing
M.~R. Albrecht, B.~R. Curtis, A.~Deo, A.~Davidson, R.~Player, E.~W.
  Postlethwaite, F.~Virdia, and T.~Wunderer, ``Estimate all the \{LWE, NTRU\}
  schemes!'' in \emph{Security and Cryptography for Networks - 11th
  International Conference, {SCN} 2018, Amalfi, Italy, September 5-7, 2018},
  pp. 351--367, 2018.
\BIBentrySTDinterwordspacing

\bibitem{sawyer2005resampling}
\BIBentryALTinterwordspacing
S.~Sawyer. (2005). Resampling data: Using a statistical jackknife. [Online].
  Available: \url{https://www.math.wustl.edu/~sawyer/handouts/Jackknife.pdf}
\BIBentrySTDinterwordspacing

\bibitem{tit/ConwayS85}
\BIBentryALTinterwordspacing
J.~H. Conway and N.~J.~A. Sloane, ``A lower bound on the average error of
  vector quantizers,'' \emph{{IEEE} Trans. Inf. Theory}, vol.~31, no.~1, pp.
  106--109, 1985.
\BIBentrySTDinterwordspacing

\bibitem{tit/Zador82}
\BIBentryALTinterwordspacing
P.~L. Zador, ``Asymptotic quantization error of continuous signals and the
  quantization dimension,'' \emph{{IEEE} Trans. Inf. Theory}, vol.~28, no.~2,
  pp. 139--148, 1982.
\BIBentrySTDinterwordspacing

\end{thebibliography}

\begin{IEEEbiography}[{\includegraphics[width=1in,height=1.25in,clip,keepaspectratio]{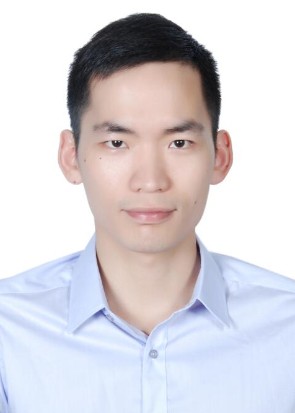}}]{Shanxiang Lyu}
	received the B.S. and M.S. degrees in electronic and information engineering from
	South China University of Technology, Guangzhou, China, in 2011
	and 2014, respectively, and the Ph.D. degree from the
	Electrical and Electronic Engineering Department, Imperial College London, UK,
	in 2018. He is currently an associate professor with the College of Cyber Security, Jinan University, Guangzhou, China. He is the recipient of the 2021 CIE Information Theory Society Yong-star Award, and the 2020 superstar supervisor award of the National Crypto-Math Challenge of China. He was in the organizing committee of Inscrypt 2020. His research interests include lattice codes, wireless communications, and cryptography.
\end{IEEEbiography}

\begin{IEEEbiography}[{\includegraphics[width=1in,height=1.25in,clip,keepaspectratio]{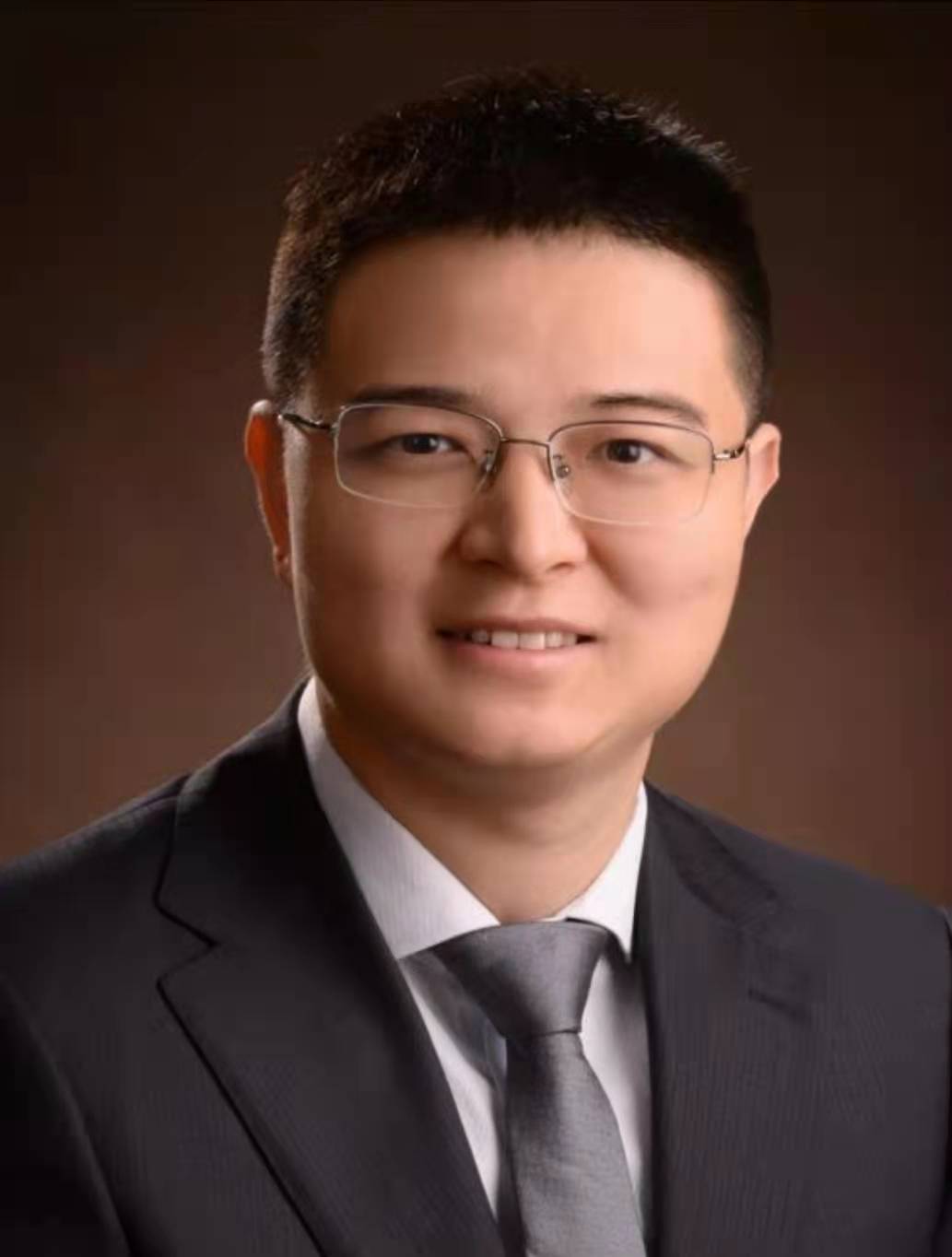}}]{Zheng Wang}
	(Member, IEEE) received the B.S. degree in electronic and information engineering from Nanjing University
	of Aeronautics and Astronautics, Nanjing, China, in 2009, and the M.S. degree in communications
	from University of Manchester, Manchester,
	U.K., in 2010. He received the Ph.D degree in communication engineering from Imperial College London, UK, in 2015.
	
	Since 2021, he has been an Associate Professor in the School of Information and Engineering, Southeast University, Nanjing, China. From 2015 to 2016 he served as a Research Associate at Imperial College London, UK. From 2016 to 2017 he was an senior engineer with Radio Access Network R\&D division, Huawei Technologies Co.. From 2017 to 2020 he was an Associate Professor at the College of Electronic and Information Engineering, Nanjing University of Aeronautics and Astronautics (NUAA), Nanjing, China.
	His current research interests include massive MIMO systems, machine learning and data analytics over wireless networks, and lattice theory for wireless communications.
\end{IEEEbiography}

\begin{IEEEbiography}[{\includegraphics[width=1in,height=1.25in,clip,keepaspectratio]{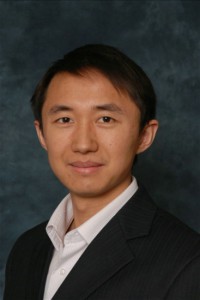}}]{Cong Ling} (S'99-M'04) 
	received the B.S. and M.S. degrees in electrical engineering from
	the Nanjing Institute of Communications Engineering, Nanjing, China, in 1995
	and 1997, respectively, and the Ph.D. degree in electrical engineering from
	the Nanyang Technological University, Singapore, in 2005.
	He had been on the faculties of the Nanjing Institute of Communications
	Engineering and King's College. He is currently a Reader (Associate Professor) with the Electrical and Electronic Engineering Department, Imperial
	College London. His research interests are coding, information theory, and
	security, with a focus on lattices.
	
	Dr. Ling has served as an Associate Editor for the IEEE TRANSACTIONS
	ON COMMUNICATIONS and the IEEE TRANSACTIONS ON VEHICULAR
	TECHNOLOGY.\end{IEEEbiography}

\begin{IEEEbiography}[{\includegraphics[width=1in,height=1.25in,clip,keepaspectratio]{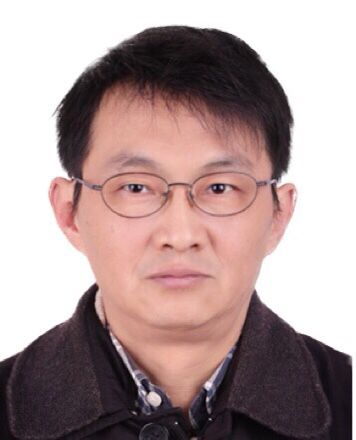}}]{Hao Chen} 
received the Ph.D. degree in mathematics from the Institute
of Mathematics, Fudan University, in 1991. He is currently a Professor of
the College of Information Science and Technology/Cyber Security, Jinan
University. His research interests include coding and cryptography, quantum
information and computation, lattices, and algebraic geometry. He has published a series of papers in Crypto, Eurocrypt, IEEE Transactions on Information Theory, etc. He was the recipient of the NSFC outstanding young scientist grant in 2002.
\end{IEEEbiography}

\end{document}